# A comprehensive review on convolutional neural network in machine fault diagnosis

Jinyang Jiao [a], Ming Zhao [a], Jing Lin [b,*], Kaixuan Liang [a]

[a] *State Key Laboratory for Manufacturing Systems Engineering, School of Mechanical Engineering, Xi'an Jiaotong University, Xi'an, Shaanxi 710049, China*

[b] *School of Reliability and Systems Engineering, Beihang University, Beijing 100083, China*

**Abstract**

With the rapid development of manufacturing industry, machine fault diagnosis has become increasingly significant to ensure safe equipment operation and production. Consequently, multifarious approaches have been explored and developed in the past years, of which intelligent algorithms develop particularly rapidly. Convolutional neural network, as a typical representative of intelligent diagnostic models, has been extensively studied and applied in recent five years, and a large amount of literature has been published in academic journals and conference proceedings. However, there has not been a systematic review to cover these studies and make a prospect for the further research. To fill in this gap, this work attempts to review and summarize the development of the Convolutional Network based Fault Diagnosis (CNFD) approaches comprehensively. Generally, a typical CNFD framework is composed of the following steps, namely, data collection, model construction, and feature learning and decision making, thus this paper is organized by following this stream. Firstly, data collection process is described, in which several popular datasets are introduced. Then, the fundamental theory from the basic convolutional neural network to its variants is elaborated. After that, the applications of CNFD are reviewed in terms of three mainstream directions, i.e. classification, prediction and transfer diagnosis. Finally, conclusions and prospects are presented to point out the characteristics of current development, facing challenges and future trends. Last but not least, it is expected that this work would provide convenience and inspire further exploration for researchers in this field.

* Corresponding author

*E-mail address*: linjing@buaa.edu.cn

# 1. Introduction

Powered by the integration innovation of intelligent manufacturing, industrial big data and industrial 4.0, the modern industry is experiencing a new revolution from the traditional manufacturing industry to intelligent industry [1, 2]. Mechanical equipment, as one of the most significant roles in this revolution, is evolving to continuously promote production and improve economic benefit. Unfortunately, various faults will inevitably be exposed during the tireless operation of machine, once the fault appears, it will cause unscheduled downtime, economic loss, even catastrophic accidents and casualties [3, 4]. However, big data generated from modern industry also affords an unprecedented opportunity to obtain an in-depth understanding of machine condition. Therefore, it is vital to seize this opportunity and advance diagnostic methods for accurate judgment and timely response on machine degradation and failure.

Over the years, a variety of approaches have been developed for machine fault diagnosis through the wisdom and efforts of researchers and engineers. Review them briefly, existing approaches can be roughly divided into four categories according to the development process, i.e. physical model-based methods, signal processing-based methods, machine learning-based methods and their hybrid [5-7]. Physical model-based methods usually require a thorough understanding for mechanisms of the machine, thus it is difficult to build accurate physics systems for modern complex mechanical equipment, especially in dynamic and noisy working environment. In addition, most of physical models are inflexible and inefficient since they are unable to be updated with real monitoring data. Different from these approaches, the signal processing-based approach aims to explore advanced signal de-nosing and filtering technologies to emphasize fault characteristic information. However, it usually requires related equipment knowledge for feature frequency calculation, moreover, the solid fault representation theory and mathematical basis are also the premise of this method. Another family named machine learning-based method, as a typical representative of data-driven approaches, has been active and brilliant with the development industrial modern industry, especially the advent of deep learning [8]. Although classical machine learning models, such as support vector machine (SVM) and $k$-nearest neighbor, have achieved remarkable progress over the past years, some drawbacks still exist when facing the higher industrial requirements [9]. For example, i) These methods generally need to extract and select features manually, which is limited in complex big data analysis. In addition, it is also difficult to effectively mine high-dimension features due to the shallow structure; ii) Feature mining and decision making are separately designed, in which the unsynchronized optimization will consume considerable time and restrict the performance; iii) With the growing diversity of sensor and complexity of machine, as well as increasing volume of data with increased dimensions and dynamics, it is difficult to obtain satisfactory diagnostic with traditional algorithms.

Deep learning, as the hottest branch of machine learning, has been witnessed the proliferation and prosperity in various fields, including image identification, speech processing and so on [10]. This is not only due to subjective factors, such as powerful capabilities of data processing, feature learning and architecture innovation, but also several external factors cannot be ignored, including i) Explosive increase of industrial big data; ii) Breakthrough of hardware, such as graphics processor unit; iii)

Stimulation from multifarious competitive task requirements. Naturally, deep learning has also raised the wave of intelligent fault diagnosis over the past five years. The popular deep learning based diagnostic models include deep auto-encoder [11], deep belief network [12], recurrent neural network [13], and convolutional neural network (CNN) [14]. Among them, convolutional network [15] has become the leading architecture and achieved state of the art performance in many benchmarks [10]. Similarity, fault diagnosis approaches using convolutional network have also developed most rapidly and a lot of research work has been published. Given the popularity of Convolutional Network based Fault Diagnosis (CNFD), a systemic review and summary is necessary to help to tease out current work and make prospects for the further research.

The CNFD framework can generally be summarized into three steps as shown in Fig. 1, including data collection, model construction, as well as feature learning and decision making. In the first step, tremendous monitoring data are collected and prepared from the concerned mechanical equipment. Next, convolutional network models are designed and constructed depend on the task requirements. Finally, the hierarchical and high-dimensional features can be adaptively learned for characterizing machinery condition. Meanwhile, the decision, such as fault classification and remaining useful life (RUL) prediction, is carried out based on the extracted features. Several merits can be clearly revealed from this framework, i) It is able to exploit the in-depth and intrinsic characteristics adaptively while alleviate the requirements of human labor as well as expert knowledge; ii) This model can flexibly update itself according to the real-time monitoring data for more practical diagnostic requirements; iii) This diagnostic framework integrates the feature extraction and decision making together and constructs an end-to-end intelligent diagnostic model.

Prior to our work, there are also several excellent review articles in machine fault diagnosis. For instance, Liu et al. [16] summarized five artificial intelligent algorithms for fault diagnosis of rotating machinery. However, their work mainly focused on the traditional machine learning models, and the review on deep learning based methods is insufficient, especially for convolutional network. Zhao et al. [17] presented a work to review several deep learning models and their applications to machine health monitoring. Although the convolutional network has also been described in their work, it was treated equally with other models and the review about CNN was not enough comprehensive. Hoang et al. [18] presented a survey on deep learning based bearing fault diagnosis, in which the literature only referred to bearing applications. Meanwhile, the summarization about CNN based methods is also incomplete. Furthermore, recent budding studies that integrate the convolutional neural networks with transfer learning technologies have not been mentioned in these papers, while these methods have gradually attracted attention since they are suitable for more practical industrial scenarios. With this in mind, this paper intends to review fault diagnosis algorithms by leveraging convolutional networks more comprehensively, meanwhile, to provide a reference for those who want to understand and promote the development of CNN technologies for fault diagnosis of machinery.

The rest of this paper is organized as follows. According the line of the Fig. 1, data collection process and several popular public datasets are described in Second 2. After that, the concept and theory of CNN and its variants are introduced in Section 3. In Section 4, the applications of convolutional network on machine fault diagnosis are comprehensively reviewed. In Section 5, some conclusions are drawn based on above review. Finally, prospects are summarized in Section 6.

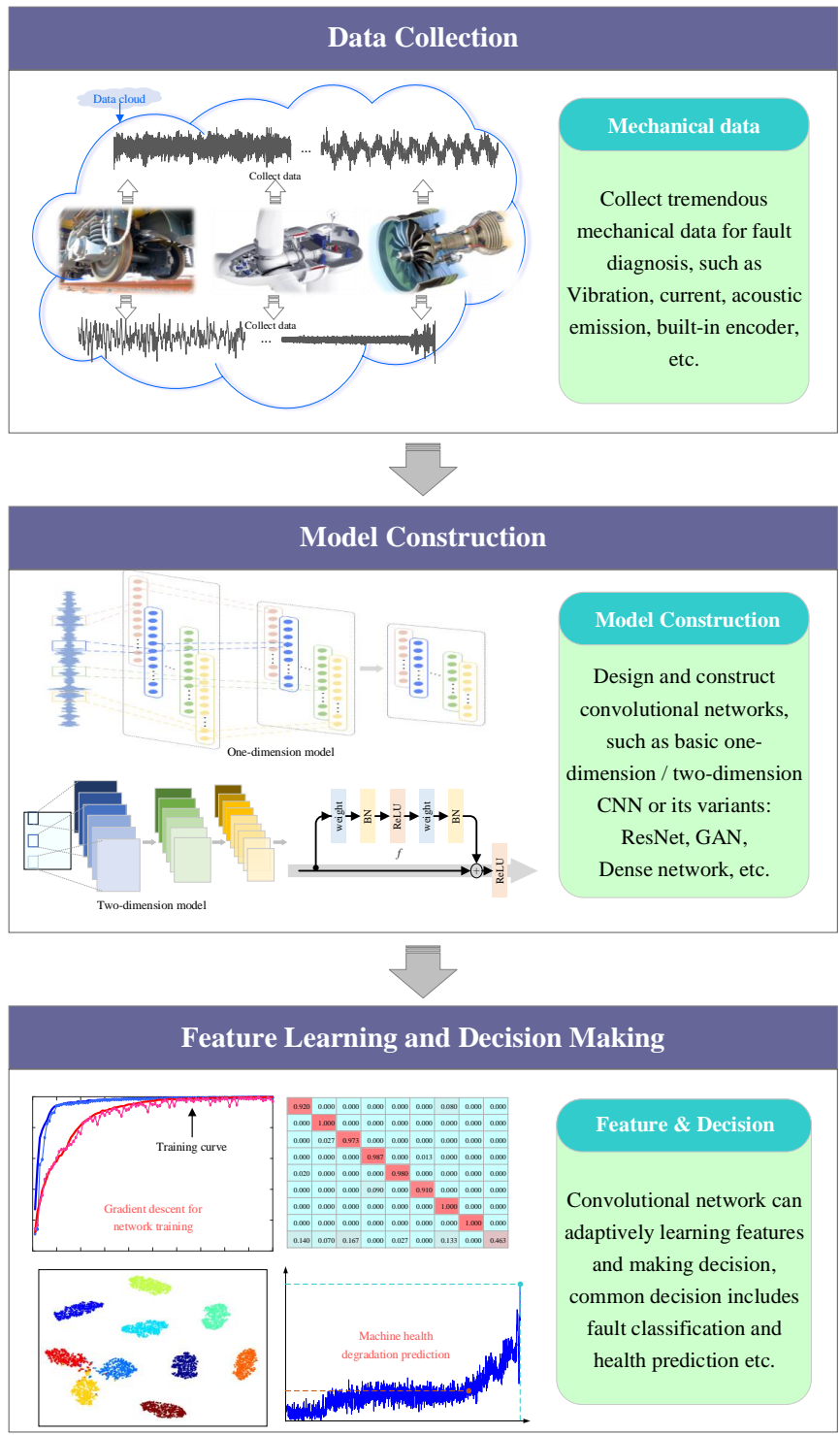

Fig. 1. The framework of general CNFD method.

**2. Data preparation**

As shown in Fig. 1, high-quality data is the premise and foundation for successfully training convolutional neural networks. In brief, there are two steps to acquire mechanical data, including sensor selection and layout as well as data sampling and storage. With the development of the sensor technology, various sensors have been applied to mechanical condition monitoring, such as vibration, current, built-in encoder [19], etc. Based on these sensors, comprehensive monitoring data can be captured for machine fault diagnosis. Among them, vibration analysis has become the most popular monitoring manner and been developed most rapidly over past years. Although these vibration-based approaches have made impressive progress, there are still some restrictions for collecting vibration data in practical industry. For instance, vibration data is often plagued by the interference of transmission path and environment noise, thus the signal-to-noise ratio of data is low. In addition, vibration data is not sensitive to low frequency response, thus it is not suitable for the condition monitoring of low speed machinery. Furthermore, the vibration sensor cannot even be installed in high temperature, high pressure or closed working environments. However, these drawbacks can be circumvented by using other sensors, for example, infrared imaging can provide a non-contact measurement method and built-in encoder signal has better signal-to-noise ratio and low frequency response. Therefore, it is of significance to comprehensively consider multiple factors, such as equipment type, working environment, monitoring object and operating condition, for selecting well-suited sensors. Next, the layout of sensor is also an important consideration since proper location can perceive much more health information and reduce the influence of transmission path and interference. Following this step, data sampling can be carried out using the data acquisition system and then data are stored by the hard disk or cloud platform for further analysis and use.

Although the process of data collection is clear and intuitive, there are still difficulties for acquiring high-quality data in real industrial scenarios [20]. For example, i) the fault data is hard to be acquired than health data since the machine is usually not allowed to run in fault condition; ii) The obtaining of life-cycle data is time consuming, expensive and even prohibitive since machine generally has a long running time from the health to failure. Fortunately, a few institutions have published datasets for public study and application. Therefore, several public datasets are introduced in following subsections, which aims to offer a guideline for researchers and engineers who intend to select these data for the evaluation of their approaches.

**2.1. CWRU bearing fault dataset**

**2.1.1. Description of the dataset**

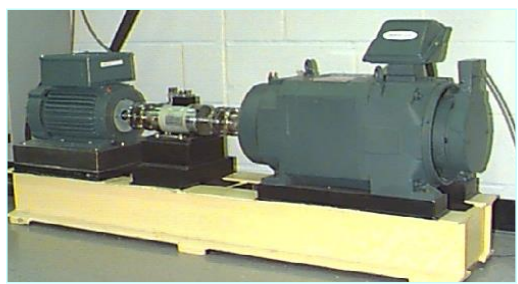

Fig. 2. Experimental platform of CWRU.

The Case Western Reserve University (CWRU) [21] bearing dataset has become one of the most popular datasets for machine fault diagnosis since it was made public. The experimental rig is shown as Fig. 2, which is consisted of an electric motor, a torque transducer/encoder, a dynamometer, and control electronics. The single point motor bearing faults simulated by the electro-discharge machining were tested in this platform, including inner race fault, outer race fault and ball fault. Each fault has different fault sizes, i.e. 7 mils, 14 mils, 21 mils, 28 mils, and 40 mils (1mil=0.001 inches). The accelerometers attached to the drive end and fan end of the motor housing were used to collect vibration data with respective sampling frequencies, i.e. 12 kHz and 48 kHz. In addition, there are four operating conditions in this dataset, including 0 hp/1797 rpm, 1 hp/1772 rpm, 2 hp/1750 rpm, and 3 hp/1730 rpm.

**2.1.2. Characteristics of the dataset**

1) The faults of this dataset were processed by the electro-discharge machining and have certain differences from the real natural industrial scenarios. Therefore, when using this dataset to construct the classification task with multi-fault, many faults are easy to be detected and the results may be lead to blind faith.

2) This dataset was collected from different sensor positions, which thus can be used to study the generalization capability of model to different sensor data.

3) There are different sizes for the same fault condition in this dataset. Thus this dataset can be applied to the transfer diagnosis scenario where the model is trained by one fault size and tested using other fault sizes.

4) This dataset includes four different working conditions. Therefore, it is suitable for the transfer fault diagnosis study, in which the training data and the test data are from different operating conditions.

**2.2. PHM09 gearbox fault dataset**

**2.2.1. Description of the dataset**

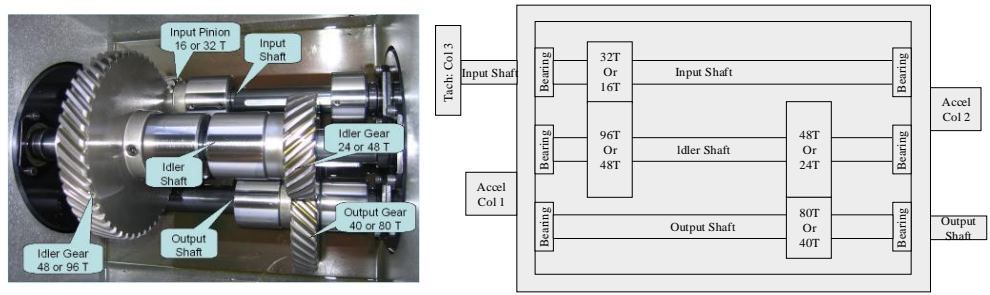

Fig. 3. Experimental rig of PHM 2009.

This database was shared by the IEEE International Conference on prognostics and health management (PHM) 2009 [22]. The schematic of the experimental rig is shown in Fig. 3, where the tested industrial gearbox contains three shafts, four gears and six bearings. Two types of gears, i.e. spur gear and spiral cut (helical) gear, were used for experimental test. The spur gear dataset contains eight health conditions and helical gear dataset has six health conditions as described in Table 1 and Table 2. In this experiment, the vibration data were sampled with 66.67 kHz sampling frequency by two accelerometers mounted on both the input and output shaft retaining plates. Meanwhile, the tachometer signals were collected by 10 pulse per revolution. In this dataset, therefore, each data file contains three columns and the first two columns are vibration data and the third column is tachometer data. Moreover, the experimental operating conditions involve five speeds and two loads, i.e. 30 Hz, 35 Hz, 40 Hz, 45 Hz, 50 Hz shaft speed and high and low loading.

**2.2.2. Characteristics of the dataset**

1) This dataset was collected from a generic industrial gearbox, in which the spur gear contains eight different health conditions and helical gear includes six different health conditions. Thus it can be used to construct the multi-classification diagnosis scenario.

2) Two accelerometers are used to synchronously sample vibration data from different positions, therefore, it is applicable for the research of the double-sensor information fusion or transfer diagnosis between sensors.

3) There are multiple hybrid faults of gears, bearings and shafts in this database, thus it is a typical case for the study of hybrid fault diagnosis.

4) This dataset includes multiple working conditions. Therefore, it can be applied to the transfer diagnosis research under different speeds and loads.

Table 1. Description of spur gear health conditions in PHM 2009 dataset.

|  | Gear | | | | Bearing | | | | | | Shaft | |
| --- | --- | --- | --- | --- | --- | --- | --- | --- | --- | --- | --- | --- |
|  | 32T | 96T | 48T | 80T | IS:IS | ID:IS | OS:IS | IS:OS | ID:OS | OS:OS | Input | Output |
| Spur 1 | G | G | G | G | G | G | G | G | G | G | G | G |
| Spur 2 | C | G | E | G | G | G | G | G | G | G | G | G |
| Spur 3 | G | G | E | G | G | G | G | G | G | G | G | G |
| Spur 4 | G | G | E | Br | B | G | G | G | G | G | G | G |
| Spur 5 | C | G | E | Br | In | B | O | G | G | G | G | G |
| Spur 6 | G | G | G | Br | In | B | O | G | G | G | Im | G |
| Spur 7 | G | G | G | G | In | G | G | G | G | G | G | KS |
| Spur 8 | G | G | G | G | G | B | O | G | G | G | Im | G |

IS: Input Shaft; ID: Idler Shaft; OS: Output Shaft; :IS: Input Side; :OS: Output Side; G: Good; C: Chipped; E: Eccentric; Br: Broken; B: Ball; In: Inner; O: Outer; Im: Imbalance; KS: Keyway Sheared.

Table 2. Description of helical gear health conditions in PHM 2009 dataset.

|  | Gear | | | | Bearing | | | | | | Shaft | |
| --- | --- | --- | --- | --- | --- | --- | --- | --- | --- | --- | --- | --- |
|  | 16T | 48T | 24T | 40T | IS:IS | ID:IS | OS:IS | IS:OS | ID:OS | OS:OS | Input | Output |
| Hel 1 | G | G | G | G | G | G | G | G | G | G | G | G |
| Hel 2 | G | G | C | G | G | G | G | G | G | G | G | G |
| Hel 3 | G | G | Br | G | G | G | G | Co | In | G | BS | G |
| Hel 4 | G | G | G | G | G | G | G | Co | B | G | Im | G |
| Hel 5 | G | G | Br | G | G | G | G | G | In | G | G | G |
| Hel 6 | G | G | G | Br | In | B | O | G | G | G | BS | G |

Hel: Helical; Co: Combination; BS: Bent Shaft.

## 2.3. Paderborn dataset

### 2.3.1. Description of the dataset

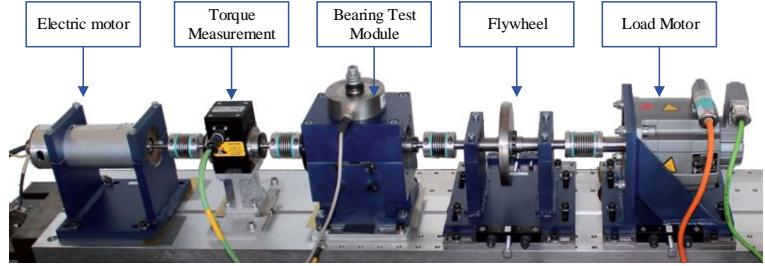

Fig. 4. Paderborn test rig for condition monitoring of rolling bearings.

The Paderborn dataset [23] was obtained using the test bench as shown in Fig. 4 and this rig was composed of an electric motor, a torque measurement shaft, a rolling bearing test module, a flywheel and a load motor. The bearings of different states were installed in the bearing test module to acquire experimental data. In total, experiments with 26 faulty bearings and 6 healthy bearings were performed, in which the fault contains 12 artificial damages as shown in Table 3 and Fig. 5 (a) and 14 real damages as shown in Table 4 and Fig. 5 (b). The motor current and vibration signals of bearing housing were synchronously measured with a sampling rate of 64 kHz. Moreover, this test rig was respectively operated under four different operating conditions as shown in Table 5 by changing the rotational speed, load torque and radial force.

Table 3. Description of test bearing with artificial damage.

| | Artificial Damage | | | | | | | | | | | |
|---|---|---|---|---|---|---|---|---|---|---|---|---|
| BC | OR | OR | OR | OR | OR | OR | OR | IR | IR | IR | IR | IR |
| ED | 1 | 2 | 1 | 2 | 1 | 2 | 2 | 1 | 1 | 1 | 2 | 2 |
| DM | EDM | EE | EE | EE | D | D | D | EDM | EE | EE | EE | EE |

BE: Bearing Component; ED: Extent of Damage; DM: Damage Method; OR: Outer Ring; IR: Inner Ring; EDM: Electric Discharge Machining; D: Drilling; EE: Manual Electric Engraving.

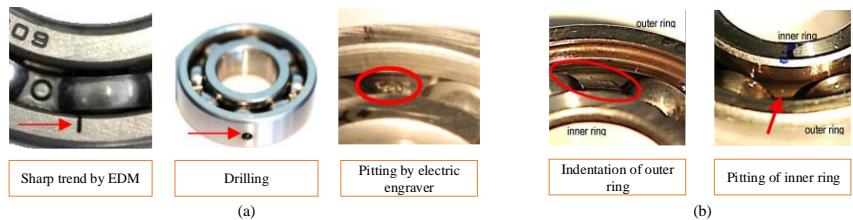

Fig. 5. Several examples of bearing damages. (a) artificial; (b) real.

Table 4. Description of test bearing with real damage caused by accelerated lifetime test.

| | Real Damage | | | | | | | | | | | | |
|---|---|---|---|---|---|---|---|---|---|---|---|---|---|
| D | FP | PI | FP | FP | PI | FP | FP | PI | FP | FP | FP | FP | FP |
| BC | OR | OR | OR | OR | OR | IR(+OR) | IR(+OR) | IR+OR | IR | IR | IR | IR | IR |
| Co | S | S | R | S | R | M | M | M | M | M | S | R | S |
| A | nr | nr | r | nr | r | r | nr | r | nr | nr | r | nr | nr |
| ED | 1 | 1 | 2 | 1 | 1 | 2 | 3 | 1 | 1 | 1 | 3 | 2 | 1 |
| CD | SP | SP | SP | SP | Di | SP | Di | Di | SP | SP | SP | SP | SP |

D: Damage (main mode and symptom); BC: Bearing Component; Co: Combination; A: Arrangement; ED: Extent of Damage; CD:

Characteristic of damage; FP: fatigue: pitting; PI: Plastic deformation: Indentations; OR: Outer Ring; IR: Inner Ring; S: Single Damage; R: Repetitive Damage; M: Multiple Damage; nr: no repetition; r: random; SP: Single Point; Di: Distributed.

Table 5. Four operating conditions.

| No. | 0 | 1 | 2 | 3 |
| --- | --- | --- | --- | --- |
| Rotational Speed (rpm) | 1500 | 900 | 1500 | 1500 |
| Load Torque (Nm) | 0.7 | 0.7 | 0.1 | 0.7 |
| Radial Force (N) | 1000 | 1000 | 1000 | 400 |

**2.3.2. Characteristics of the dataset**

1) This dataset contains different damage states and thus can be used for the multi-fault classification study. Besides, this dataset is more comprehensive than CWRU bearing data since it takes into account the artificial and realistic bearing damages simultaneously.

2) In this dataset, the motor current and vibration signals were synchronously sampled for bearing health information collection. Therefore, on one hand, the motor current signal or the vibration signal can be independently used for bearing diagnosis and comparison study; on the other hand, multi-sensors information fusion based diagnosis can be studied using this dataset.

3) This dataset can be used for the transfer diagnosis according to different fault formation mode, that is, the model is trained using the artificial fault data and tested using the real fault data.

4) There are four operating conditions with different speeds, load torques and radial forces. Thus this dataset is suitable to the transfer diagnosis study under different working conditions.

**2.4. IMS bearing dataset**

**2.4.1. Description of the dataset**

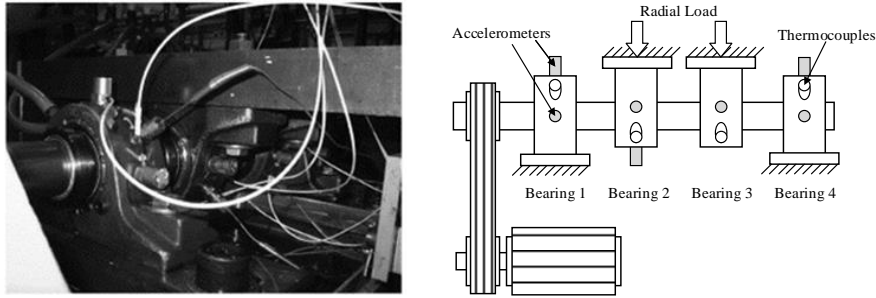

Fig. 6. The IMS bearing test rig.

This bearing dataset was established by the center for Intelligent Maintenance Systems (IMS) of University of Cincinnati [24] and the used test rig is presented in Fig. 6. In this test rig, four Rexnord ZA-2115 double row bearings were installed on the shaft for testing. The rotation speed was kept constant at 2000 RPM by an AC motor coupled to the shaft via rub belts. Besides, a radial load of 6000 LBS was applied to the shaft and bearing by a spring mechanism. The vibration data were acquired by accelerometers attached on the bearing housings with the sampling rate of 20 kHz. In total, there are three experiments and each described a test-to-failure task as shown in Table 6. The inner race defect occurred in bearing 3 and roller element defect in bearing 4 at the end of the first experiment. In the second and third experiment, the outer race failure finally occurred in bearing 1 and bearing 3, respectively.

Table 6. Description of bearing conditions in three experiments.

| Experiment | Bearing 1 | Bearing 2 | Bearing 3 | Bearing 4 |
| --- | --- | --- | --- | --- |
| 1 | UD | UD | IRD | RED |
| 2 | ORD | UD | UD | UD |
| 3 | UD | UD | ORD | UD |

UD: undamaged; IRD: inner race damage; RED: roller element damage; ORD: outer race damage.

**2.4.2. Characteristics of the dataset**

1) This dataset contains four different health conditions, i.e. health, roller fault, outer race fault and inner race fault, which can be used to study the issue of bearing classification.

2) Each data described a run-to-failure experiment, thus researchers can employ this dataset to study the bearing RUL prediction.

3) The bearings of this experiment experienced an "increase-decrease-increase" degradation trend, in which the reason of "decrease" is the self-healing nature of the damage. As a result, selecting data during this period will increase the difficulty of fault diagnosis.

4) There is only one operating condition in this dataset, which limits the diversity of the data. In addition, the lifetime of each unit has distinct discrepancies, which increases the difficulty of RUL prediction.

**2.5. C-MAPSS dataset**

**2.5.1. Description of the dataset**

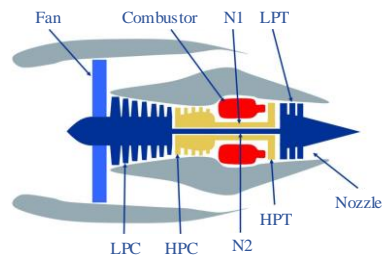

Fig. 7. The diagram of the simulated engine.

C-MAPSS dataset [25] was provided by Commercial Modular Aero-Propulsion System Simulation (C-MAPSS) to simulate the turbofan engine degradation. The diagram of simulated engine is shown in Fig. 7. This system has 14 inputs, i.e. the fuel flow and a set of 13 health parameters, which allows the user to simulate the effects of faults and deterioration in any of the engine's five rotating components. There are 58 outputs including various sensor responses and operability margins, in which a total of 21 sensor variables were used to measure the health states of the engine. In total, five subsets are included in this dataset as listed in Table 7 and each trajectory has a specific initial wear level and degradation process.

Table 7. Five subsets included in the turbofan engine degradation dataset.

| No. | Train trajectories | Test trajectories | Conditions | Fault modes |
|---|---|---|---|---|
| 1 | 218 | 218 | -- | -- |
| 2 | 100 | 100 | One (sea level) | HPC Degradation |
| 3 | 260 | 259 | Six | HPC Degradation |
| 4 | 100 | 100 | One (sea level) | HPC and Fan Degradation |
| 5 | 248 | 249 | Six | HPC and Fan Degradation |

**2.5.2. Characteristics of the dataset**

1) This dataset was generated from the simulation software and thus it has a certain difference with the experimental data or realistic industrial scenario.

2) There are sufficient training samples in this dataset, which thus is able to train the complex convolutional networks for prediction.

3) This dataset contains 21 different observation features, such as the temperature, pressure, speed, etc., which means that this dataset can be applied for the study of multi-sensor information fusion.

4) There are different operating conditions for the same fault mode. Thus, it can be used to evaluate the generalization capability of model.

**2.6. PHM 10 CNC Milling Machine Cutters Dataset**

**2.6.1. Description of the dataset**

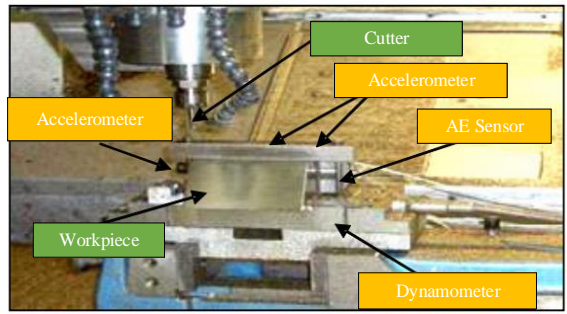

Fig. 8. The high-speed CNC milling machine cutters.

This dataset was shared in the 2010 PHM Society Conference Data Challenge [26], which focused on RUL estimation of a high-speed CNC milling machine cutters. The experimental rig is shown in Fig. 8 and the operation parameters are listed in Table 8. A Kistler quartz 3-component platform dynamometer was mounted between the work piece and machining table to measure the cutting forces, three Kislter accelerometers were mounted on the work piece to measure the machine tool vibrations of cutting process in X, Y, Z direction, respectively, and a Kistler acoustic emission sensor was mounted on the workpiece to monitor the high frequency stress wave generated by the cutting process. There are six individual cutter data in this dataset in total.

Table 8. The description of operation parameters.

| Description | Value |
| --- | --- |
| Running speed of the spindle | 10400 rpm |
| Feed rate in the $x$ direction | 1555 mm/min |
| Depth of cut in the $y$ direction | 0.125 mm |
| Depth of cut in the $z$ direction | 0.2 mm |
| Sampling frequency | 50 kHz |

**2.6.2. Characteristics of the dataset**

1) This dataset was conducted under the dry milling environment, therefore, it has certain differences with the real milling process. However, real milling data are quite difficult to be acquired due to the cost or commercial competition, this dataset is still a good choice for the study of RUL prediction of milling machine cutters.

2) Each data file is composed of three-dimension cutting force data, three-dimension vibration data and acoustic emission signal. Therefore, it can be used to explore single-sensor or multi-sensor fusion based prediction scenarios.

3) Although this dataset has six cutter data, there are only three cutters are labeled. Thus the amount of data may be insufficient for building complex diagnostic networks.

4) This experiment was performed in one milling operating condition, which limits the diversity of data and restricts the construction of cross-validation prediction scenarios.

**2.7. FEMTO dataset**

**2.7.1. Description of the dataset**

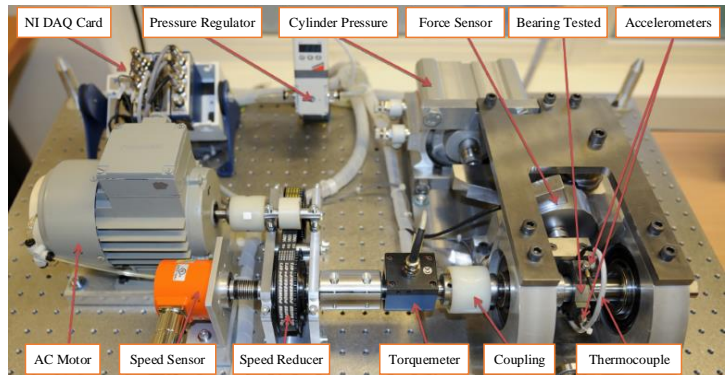

Fig. 9. PRONOSTIA experimental platform.

FEMTO dataset [27] was acquired from the PRONOSTIA experimental platform designed by Franche-Comté Electronics, Mechanics, Thermal Processing, Optics-Sciences and Technologies institute (FEMTO), which aimed to provide the experimental data to characterize the degradation of bearings. This dataset was also used for the prognostic challenge in the IEEE International Conference on PHM 2012. The overview of this test rig is presented in Fig. 9, which is composed of a rotating part (the asynchronous motor with a gearbox and two shafts), a degradation generation part (with a radial force applied on the tested bearing) and a measurement part (sensors). All bearings are healthy and not seeded with any defects at the beginning of the test. Two types of sensor, i.e. thermocouple and accelerometers (horizontal and vertical) were used to collect the temperatures and vibration signals of the testing bearing, the sampling frequencies of vibration and temperature were set as 25.6 kHz and 10 Hz, respectively. The bearing life is considered terminated when the amplitude of the vibration signal exceeds 20 g, which aims to avoid propagation of damages to the whole test bed. In summary, this dataset contains 17 start-to-end data of bearings.

**2.7.2. Characteristics of the dataset**

1) This is a full-life bearing dataset with real damages, thus it is a good choice for the study of bearing RUL prediction. However, the train set is small while the spread of the life duration of all bearing is wide, which means that the RUL estimation is more challenging using this dataset.

2) Although two-direction vibration signals are collected, the vertical vibration signals provide less useful information than the horizontal ones for tracking the bearing degradations according to the related literature [28, 29].

3) This dataset provides a natural degradation process since the bearings are healthy and not seeded with any defects at the beginning of the tests. But this dataset presents no prior information about the properties of the damages.

4) The degradation and fault patterns are discriminative for distinct bearings even under the same operating condition due to various factors, which thus increases the difficulty of bearing RUL estimation.

**2.8. Epilog**

Table 9. The summary of different datasets.

| Name | Monitoring object | Multi-sensors | Classification | Prediction | Transfer diagnosis |
|---|---|---|---|---|---|
| CWRU | Motor Bearing | √ | √ | × | √ |
| PHM 09 | Gearbox, Baring, Shaft | √ | √ | × | √ |
| Paderborn | Bearing | √ | √ | × | √ |
| IMS | Bearing | × | √ | √ | × |
| C-MAPSS | Turbofan engine | √ | × | √ | √ |
| PHM 10 | Milling Machine Cutters | √ | × | √ | × |
| FEMTO | Bearing | √ | × | √ | √ |

In this section, data collection is firstly introduced from sensor selection and layout to data sampling and storage. To provide more convenience for relevant scholars, we summarize seven popular public datasets and list a concise summary in Table 9. This table firstly shows the main monitoring object of each dataset in second column. Then whether each dataset contains multiple sensing information is summarized. Finally, the application scenarios, including classification, RUL prediction, and transfer diagnosis are illustrated and compared in the subsequent columns. Furthermore, it is noteworthy that the datasets introduced in this work are just several popular ones and there are still some other available public datasets for the use and reference, such as MFPT fault dataset [30], XJTU-SY bearing dataset [31], University of Connecticut gear fault dataset [32]. In addition, the international conferences on PHM from the PHM society or the IEEE reliability society often provide some valuable datasets for researchers and engineers.

## 3. Convolutional neural network and its variants

Convolutional neural network, as the leader of deep learning models, has become a milestone technique and achieved state-of-the-art performances in various computer vision and pattern recognition tasks. Similarly, convolutional neural network has also shined in the field of machine fault diagnosis. From the perspective of completeness, the basic theory on CNN are firstly presented before reviewing its applications, which aims to provide the understanding and preparation for the researchers, engineers and even beginners who intend to apply convolutional networks for fault diagnosis.

### 3.1. The basic convolutional neural network

Take one-dimensional mechanical signal as an example, a basic convolutional neural network is displayed in Fig. 10, which contains one input layer, multiple convolution-pool and fully-connected layers and one output layer. Moreover, two popular operations, including batch normalization and dropout, are also embedded in this structure, which can help to improve the model performance. In following subsections, each operation will be introduced respectively.

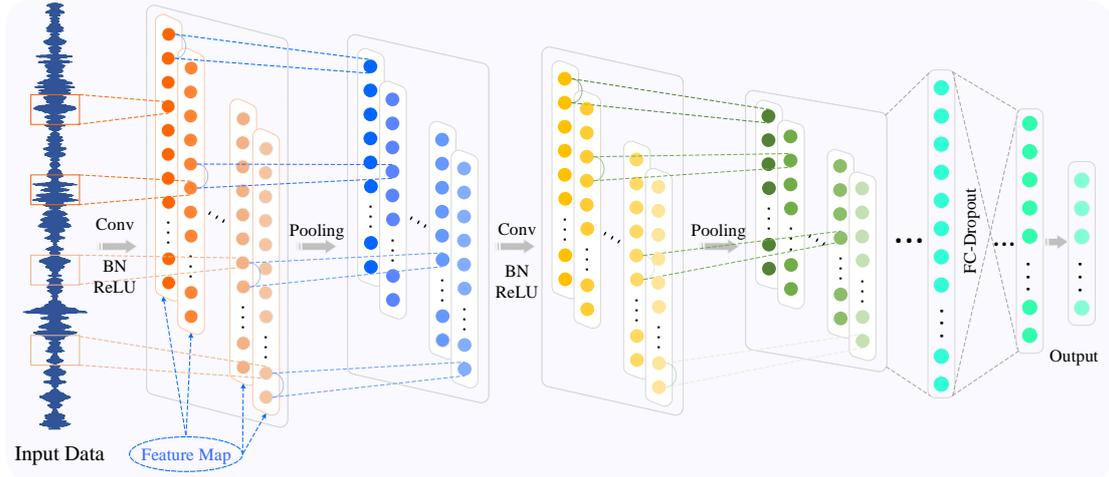

Fig. 10. An example of convolutional neural network. Conv: Convolution; BN: Batch Normalization; ReLU: Rectified Linear Unit activation function; FC: Fully-connected layer.

#### 3.1.1. Convolution and nonlinear activation

The convolutional kernel (filter) plays central role in the convolutional layer, which endows two key core ideas, i.e. sparse connection and shared weight. Each kernel is connected with the local patch in the feature maps of the previous layer, meanwhile, the weights remain unchanged when the kernel slides on these maps. In general, multiple kernels are contained in one convolutional layer, which aims to learn comprehensive feature representations. Mathematically, let $\mathbf{x} \in \mathbb{R}^d$ be the $d$-dimensional mechanical data, the $i$-th convolutional feature in the $j$-th map can be described as:

$$\mathbf{c}^j = \mathbf{x} * \mathbf{w}^j + b^j \tag{1}$$

where $\mathbf{w}^j \in \mathbb{R}^h$ represents the $j$-th filter, it is used to code the input $\mathbf{x}$ and generate the $j$-th feature map $\mathbf{c}^j=[c_1^j, c_2^j,...,c_{d-h+1}^j]$; $b^j$ denotes the bias term.

After the convolutional operation, the non-linear activation function, such as sigmoid function, tanh function, and rectified linear unit (ReLU) [33] as shown in Fig. 11, is usually employed to achieve the feature transformation. At present, ReLU is widely used in CNN since it not only computes much faster than sigmoid and tanh function, but also can alleviate the issue of gradient vanishing. However, a potential disadvantage of ReLU unit is that it has zero gradient whenever the unit is not active. This may cause units that do not active initially never active as the gradient-based optimization will not adjust their weights. To alleviate this problem, more advanced activation function has been proposed, such leaky ReLU [34].

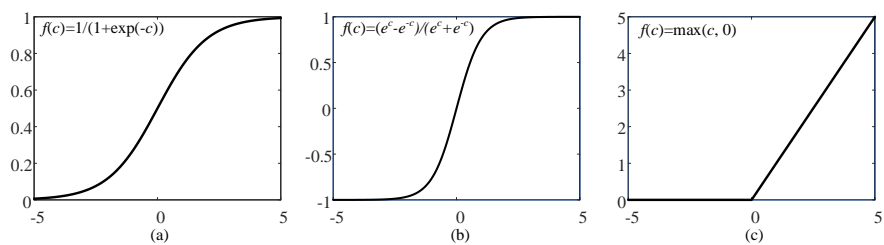

Fig. 11. Three nonlinear activation functions. (a) sigmoid; (b) tanh; (c) ReLU.

### 3.1.2. Pooling

The pooling layer, as another important operation in CNN, aims to reduce the dimension of features and enable features more robust. The common pooling operations include the max pooling and the average pooling and the difference between them is whether to take the maximum or average value in the pooling region [35]. Take the max pooling as an example, the mathematical description is given as follows:

$$po_k^j = \max\{c_{k:k+p-1}^j\} \qquad (2)$$

where $c_{k:k+p-1}^j$ represents input and $p$ is the pool size; $po_k^j$ denotes the maximum value in the corresponding pooling region.

### 3.1.3. Batch normalization

The batch normalization (BN) [36] has become a popular technique to alleviate internal covariance shift and promote network training. Mathematically, given the $d$-dimensional input $\mathbf{x}=\{x^{(1)},...,x^{(d)}\}$, the operation of BN is described as follows:

$$\tilde{x}^{(k)} = \frac{x^{(k)} - \mathbb{E}[x^{(k)}]}{\sqrt{\text{Var}[x^{(k)}]}}, h^{(k)} = \gamma^{(k)}\tilde{x}^{(k)} + \beta^{(k)} \qquad (3)$$

where $x^{(k)}$ and $h^{(k)}$ represent the $k$-th activation input and output, respectively; $\mathbb{E}(\cdot)$ and $\text{Var}(\cdot)$ denote the expectation and variance; $\gamma^{(k)}$ and $\beta^{(k)}$ stand for the parameters to be learned.

### 3.1.4. Dropout

Dropout [37] is a technique that prevents overfitting and provides a way of approximately combining different networks. The key operation of dropout is to randomly drop neuron units (along with their connections) of the network during training as shown in Fig. 12. Specifically, a unit is present with probability $p$ at training time, while the unit is always present and the weights are multiplied by $p$ at test time.

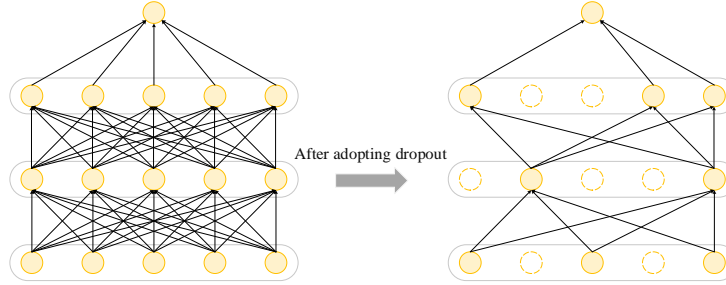

Fig. 12. The example of dropout operation.

### 3.1.5. Fully-connected layer

After stacking multiple convolution-pool modules, the fully-connected layers are usually employed to process features further. The mathematical calculation of the fully-connected layer is the same as the traditional perception, it can be described as:

$$\mathbf{fc}^l = \sigma(\mathbf{w} \cdot \mathbf{fc}^{l-1} + b) \tag{4}$$

where $\mathbf{fc}^l$ represents the output features of $l$-th fully-connected layer; $\mathbf{w}$ and $b$ stand for the connection weight and bias, respectively; $\sigma(\cdot)$ denotes the non-linear activation function.

### 3.1.6. Decision layer

After the feature extraction, the decision layer is usually followed to get the final results. There are usually two typical outputs in fault diagnosis problem, the one is classification label output and the other is single variable output, such as RUL prediction. Softmax function has become one of the most popular choices for the classification task owing to its convenience and effectiveness. Suppose the input data $\mathbf{x}$ that belongs to one of the class $N_c$, then the Softmax output that estimates the category probability of $\mathbf{x}$ can be calculated as:

$$\text{Softmax}(\mathbf{x}) = \begin{bmatrix} p(y=1) \\ p(y=2) \\ ... \\ p(y=N_c) \end{bmatrix} = \frac{1}{\sum_{j=1}^{N_c} e^{\theta_j^T \mathbf{x}}} \begin{bmatrix} e^{\theta_1^T \mathbf{x}} \\ e^{\theta_2^T \mathbf{x}} \\ ... \\ e^{\theta_{N_c}^T \mathbf{x}} \end{bmatrix} \tag{5}$$

where $\boldsymbol{\theta}=[\theta_1, \theta_2, ..., \theta_{N_c}]^T$ stands for the parameters. Note that the value of Softmax(x) is positive and the sum of each item is 1.

### 3.2. Variants and extension of convolutional network

In addition to the original convolutional network, the improved variants have also been developed and applied to the field of fault diagnosis for more excellent performance. Therefore, several common variants will be introduced briefly in this subsections, including residual network, densely connected convolutional network, and generative adversarial convolutional network.

#### 3.2.1. Residual network

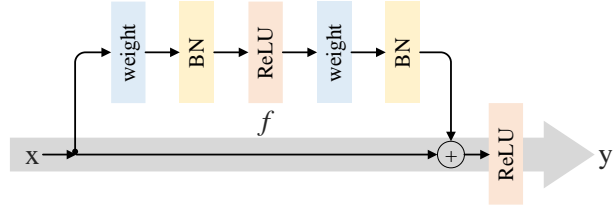

Fig. 13. The residual learning connection, where weight represents convolution; BN stands for batch normalization; and ReLU is rectified linear unit.

Stacking deeper convolutional layers in the regular CNN blindly will occur the performance degradation or gradient vanishing/exploding problem. To address these problems, a novel convolutional model, named residual network (ResNet) [38], is proposed and has become the typical representative in deep networks owing to noticeable improvements. The ResNet is generally composed of multiple residual learning blocks and each contains the convolutional layer, BN layer and activation layer as shown in Fig. 13. From this figure, it can be seen that the output of the residual block is calculated as $\mathbf{y} = f(\mathbf{x}, \mathbf{w}) + \mathbf{x}$, where f denotes the residual mapping to be learned; $\mathbf{w}$ represents the parameters. The operation $f + \mathbf{x}$ is carried out by a shortcut connection of element-wise addition. Note that a projection by shortcut connection is performed to match the input and output when they have different dimensions.

#### 3.2.2. Densely connected convolutional network

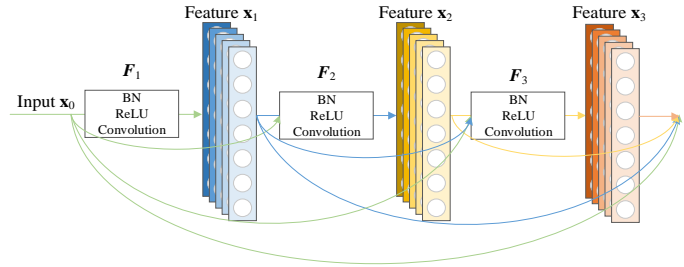

Fig. 14. The dense connection block.

In addition to ResNet, another popular deeper structure named densely connected convolutional network (DenseNet) [39] has also attracted increasing attentions owing to its excellent performance. The DenseNet is composed multiple dense connection block, in which the current layer receives the feature maps from all previous layers as shown in Fig. 14. More specifically, given the input $\mathbf{x}_0$, the feature calculation of the $l$-th layer can be described as $\mathbf{x}_l = F_l([\mathbf{x}_0, \mathbf{x}_1, ..., \mathbf{x}_{l-1}])$, where $[\mathbf{x}_0, \mathbf{x}_1, ..., \mathbf{x}_{l-1}]$ denotes the concatenation of the feature maps generated in layer 0,…,$l$-1; $F_l$ represents a composite function of three consecutive operations, i.e. BN, ReLU, and convolution.

**3.2.3. Generative adversarial convolutional network**

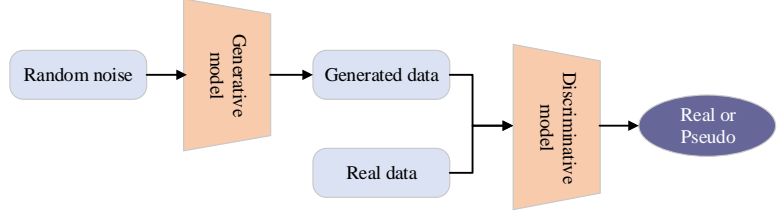

Fig. 15. The framework of GAN.

Generative adversarial convolutional network (GAN) [40] has become an important research hotspot with promising performance on data generation. The GAN usually contains two models, i.e. a generative model $G$ and a discriminative model $D$, which are pitted against each other to find a Nash equilibrium. As shown in Fig. 15, the $G$ is trained to learn the distribution of real data and generate samples from noise and the $D$ is trained to distinguish the real samples and pseudo samples by enabling the high output probability of real samples and the low probability of generative samples. In other words, the optimization of adversarial learning is a minimax game as follows:

$$\min_G \max_D \mathcal{L}(D,G) = \mathbb{E}_{\mathbf{x} \sim P_{\text{data}(\mathbf{x})}}[\log D(\mathbf{x})] + \mathbb{E}_{\mathbf{z} \sim P_{\mathbf{z}(\mathbf{z})}}[\log(1 - D(G(\mathbf{z})))] \quad (6)$$

where $p_{\text{data}(\mathbf{x})}$ represents true data distribution; $p_{\mathbf{z}(\mathbf{z})}$ is the distribution of random noise z.

**3.3. Epilog**

In this section, the concept on basic CNN and several variants is introduced, which aims to help readers better understand the work mechanism of convolutional network. Unfortunately, there is no specific guidelines for architecture selection and design, thus researchers need to optimized networks following own task requirements. It is necessary to pay attention to the same network architecture when designing the comparison approaches for a fair result. Furthermore, most advanced networks are developed based on image data feature, thus engineers are strongly encouraged to explore novel convolutional networks that fits the characteristic of mechanical data, which is promising for more excellent fault diagnosis.

**4. Applications on CNFD**

In this section, applications on CNFD are systematically reviewed and summarized, which covers published journal and conference papers in recent three years. In particular, we introduce these literature according to the following three aspects: fault classification, health prediction, and transfer diagnosis.

**4.1. Applications on fault classification**

Fault classification is the earliest and the widely studied sub-field in CNFD inspired directly by image classification. In this section, the applications on fault classification are systematically reviewed. To make the narrative more organized, we elaborate these literatures depend on the structure characteristics of convolutional network and categorize them from three aspects, i.e. two-dimension (2-D) convolutional network based classification, one-dimension (1-D) convolutional network based classification, and fault classification based on convolutional network variants.

**4.1.1. 2-D convolutional network based classification**

At the beginning, the convolutional networks used for machine fault diagnosis are the original 2-D structure by imitating the image processing. Since mechanical data is a 1-D time series in almost all cases, the main idea is to convert the 1-D data into the 2-D form in this case. Therefore, we firstly refine various signal conversion methods and then summarize the related applications.

- Data matrix transformation

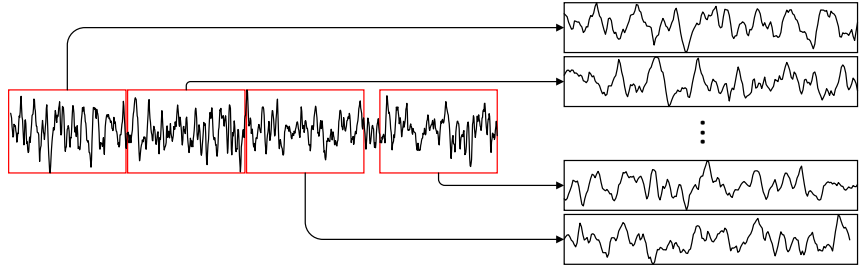

Fig. 16. Simple illustration of data matrix transformation.

Data matrix transformation refers to that researchers directly arranged raw mechanical data into a 2-D format as the model input as shown in Fig. 16. Guo et al. [41] transformed vibration data into matrix 2-D and then proposed a hierarchical CNN with adaptive learning rate for fault pattern recognition and fault size evaluation. Wang et al. [42] arranged the raw vibration signal into 2-D input and introduced an adaptive CNN for bearing fault diagnosis, in which the particle swarm optimization method was added to determine the main parameters of CNN. Shao et al. [43] proposed a bearing fault diagnosis method based on the deep convolutional belief network and the compressed sensing technique. Similarly, they [44] also utilized the auto-encoder to compress data for fault diagnosis of electric locomotive bearing. Wang et al. [45] constructed mechanical data into Hankel matrix and proposed a CNN based hidden Markov model for bearing fault classification. Gong et al. [46] firstly integrated the temporal and spatial multichannel raw signals to construct the model input. After that, the 2-D convolutional network was designed for feature learning and SVM was used for fault classification. Jing et al. [47] presented an adaptive multi-sensor data fusion based CNN method for planetary gearbox fault diagnosis, which aimed to optimize a combination of different fusion levels to satisfy the requirements of different diagnosis tasks. Chen et al. [48] fused horizontal and vertical direction vibration data into 2-D matrix and presented a deep CNN for health state identification of planetary gearboxes. Han et al. [49] presented a diagnostic framework of complex systems by combining the spatiotemporal pattern network with convolutional network, in which the former was used for spatiotemporal feature learning and the latter was used for condition classification. Yang et al. [50] used the hierarchical symbolic analysis to process original signal and then built a three-layer convolutional network for fault diagnosis of rotating machinery. Liu et al. [51] presented a dislocated time series CNN for fault classification, in which a dislocated layer is introduced to constructed 2-D input data. Yang et al. [52] transformed multi-source vibration signals to construct the 2-D matrix and then proposed a CNN based method for fault diagnosis of reciprocating compressor.

- Image transformation

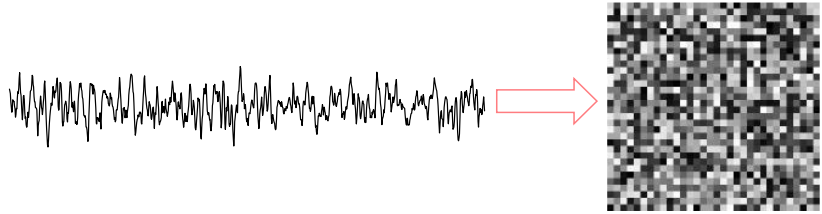

Fig. 17. Simple illustration from mechanical data to image.

Image transformation means that researchers try to convert 1-D mechanical signals into the image in pixel format as shown in Fig. 17. Xia et al. [53] presented a fault diagnosis method for rotating machinery based on multiple sensors fusion and CNN. In this method, raw vibration signals from sensors of different locations were aligned into 2-D images as the input. Hoang et al. [54] converted raw vibration signals into vibration images, then a simple two-layer convolutional model was constructed for rolling bearing fault classification. Zhang et al. [55] proposed an equitable sliding stride segmentation approach to expanse data volume whilst converted the data into images. Next a hybrid model based on the convolutional network and bi-gate recurrent unit was constructed for feature learning and classification. Hoang et al. [56] converted the motor current signals into gray images and presented a decision level fusion based CNN for bearing fault identification. Wang et al. [57] used multi-sensor data fusion to construct image data and then designed a four-layer convolutional network for fault classification. Hu et al. [58] used the compressed sensing technology to reduce data size and retain information as much as possible and transform data into image pixel. After that, an improved multi-scale convolutional network was constructed for fault recognition of machinery. Wang et al. [59] converted multi-sensor vibration signals into RGB color images to refine features and enlarge the differences between different types of fault signals. Then, an improved LeNet-5 was designed for fault diagnosis. In [60], raw mechanical signals were transformed into a square matrix through non-overlapping cutting and normalization. Then a modified LeNet-5 was designed for feature learning and fault classification.

- Time or frequency domain transformation

Another type of conversion is to use the statistics of time or frequency domain as input information of convolutional network. Chen et al. [61] calculated statistical measures of vibration signals from the time and frequency domains as the model input and then applied one-layer convolutional network for fault identification of bearings and gears. Janssens et al. [62] utilized the discrete Fourier transform to process the accelerometer signals and presented a simple convolutional network for bearing condition recognition. Bhadane et al. [63] used the statistical features extracted from vibration data as the model input and developed a 2-D CNN for bearing fault classification. Lu et al. [64] proposed a convolutional network based health state classification method for rolling bearing. In this method, the time and frequency domain features of vibration data were extracted to build the input matrix. Li et al. [65] used the root mean square maps from the spectrum of two vibration data as the input and presented a CNN with an improved Dempster-Shafer evidence theory for bearing fault diagnosis. Tra et al. [66] utilized the spectral energy maps of the acoustic emission signals as the model input. Then a CNN with the stochastic diagonal Levenberg-Marquardt algorithm was proposed for incipient bearing fault diagnosis under variable operating speeds. Prosvirin et al. [67] transformed 1-D acoustic emission signals into 2-D kurtogram images and then utilized the CNN for feature extraction and bearing fault classification. Tian et al. [68] integrated features from time and frequency domains as the input and developed a deep CNN with an immunity algorithm for rolling bearing fault diagnosis. Tra et al. [69] used the energy distribution maps of acoustic emission spectra to train a convolutional network for fault diagnosis under variable speed conditions. Li et al. [70] constructed feature images from multi central frequencies as well as vibration frequency spectrum and then used CNN to process these images for gear fault identification. Yao et al. [71] integrated features of time and frequency domains from multi-channel acoustic signals as input data and established a convolutional network for gear fault diagnosis. Kien et al. [72] visualized the spectrums of vibration signals as grayscale images and then proposed a deep convolutional network to process these images for crack detection of gears.

- Wavelet transform

Wavelet transform (WT) preprocessing method is to convert mechanical time series into 2-D time-frequency representation as the network input. Ding et al. [73] used the wavelet packet energy image as the input of deep convolutional network and presented an energy-fluctuated multiscale feature mining approach for spindle bearing fault diagnosis. Gao et al. [74] used the complex Morlet wavelet to acquire the 2-D time frequency maps from vibration signals. Then CNN was designed for rolling bearing fault diagnosis. Guo et al. [75] employed the continuous WT to decompose vibration signals into scalogram according to the rotating speed. Then a Pythagorean spatial pyramid pooling based convolutional network was presented for bearing fault diagnosis. Xu et al. [76] utilized the WT to convert vibration signals into 2-D grayscale images. Then LeNet-5 was built to learn multi-level features and the random forest classifiers were used for bearing fault classification. Islam et al. [77] employed the discrete wavelet packet transform to process the acoustic emission (AE) signals and proposed a convolutional diagnostic model for bearing fault classification. Sun et al. [78] used the dual-tree complex WT to acquire the multiscale features to train the CNN for gear fault recognition. Cabrera et al. [79] proposed a convolutional diagnostic network pre-trained by the stacked convolutional auto-encoder for fault severity assessment of helical gearbox, in which the time-frequency features acquired by the WT were used as the model input. Han et al. [80] proposed a dynamic ensemble convolutional neural network for gear fault diagnosis. In their method, wavelet packet transform was employed to construct multi-level wavelet coefficients matrices for representing the nonstationary vibration signals. Then multiple paralleled CNNs with shared parameters and a dynamic ensemble layer were designed for feature extraction and fault classification. Grezmak et al. [81] presented an explainable deep CNN with layer-wise relevance propagation for gearbox fault diagnosis, in which the time-frequency images from continuous WT were used as the model input. Liang et al. [82] adopted the WT to extract time-frequency information from vibration signals to train a CNN for compound fault diagnosis of gearbox. Guo et al. [83] used the continuous WT to convert the original mechanical signals as the input of the convolutional network for rotor fault diagnosis. Shao et al. [84] converted the vibration and current signals into the time-frequency representation by the WT. Then a deep CNN was designed to predict induction motor conditions. Hsueh et al. [85] used the empirical WT to process the current signals into grayscale images and then trained a convolutional network for induction motor fault classification. Chen et al. [86] employed the continuous WT to process raw vibration signals and then designed a convolutional model with a square pooling architecture for feature extraction. Finally, extreme learning machine was used for fault classification. Cao et al. [87] firstly adopted the dual tree wavelet to decompose the machine spindle vibration signals. Then the reconstructed sub-signal sequences from different scales and their Hilbert envelope demodulation spectra were stacked to train convolutional neural network for tool wear state identification.

- Short Time Fourier Transform

Analogously to wavelet transform, Short Time Fourier Transform (STFT), as another common time-frequency analysis approach, has also been used for data preprocessing in CNFD. Verstraete et al. [88] utilized the STFT to generate image representations of raw vibration signals and then constructed a CNN for bearing classification. Pandhare et al. [89] employed the time-frequency features obtained by STFT to train convolutional network model for bearing fault diagnosis. Xin et al. [90] used STFT to calculate TF features and then employed the sparse auto-encoder and convolutional network for feature extraction. Finally, Softmax classifier was used to obtain the classification results. Yu et al. [91] transformed the phonetic signals into spectrograms using STFT and then used VGG 16 for wind turbine fault diagnosis. In [92], Wen el al. presented a snapshot ensemble convolutional network for fault diagnosis of pump and bearing, this method can find the proper range of learning rate when facing a new dataset. In [93], Wang et al. proposed a convolutional neural network based motor fault diagnosis method, in which STFT was used to pretreat raw signals to acquire time-frequency maps.

- Other preprocessing technologies

Li et al. [94] used the S-transform to process original data into the time-frequency coefficients matrix. Then CNN was designed for feature learning and fault classification. Wen et al. [95] proposed a convolutional network based two-level hierarchical diagnosis network, in which S-transform was used to preprocess data. Jeong et al. [96] employed shaft orbit shape images as monitoring information to train the convolutional neural network for fault diagnosis. Waziralilah et al. [97] used the Gabor transform to process the raw vibration signal and then presented a CNN for bearing fault diagnosis. Zhao et al. [98] used the Hilbert transform and synchrosqueezing transform to calculate the TF representations of the vibration signals. Then these features were used to train a convolutional model for bearing fault classification. Janssens et al. [99] employed VGG to process infrared thermal video for condition detection of the machine. Jia et al. [100] presented a convolutional network based fault detection model by processing infrared thermography images. Li et al. [101] proposed a rotating machinery condition monitoring method, in which the CNN was designed to process the infrared thermal images for feature extraction and fault classification. Chen et al. [102] developed a CNN based degradation state identification approach for planetary gear, in which the singular spectrum of raw data was used to train the model. Wang et al. [103] employed the singular value decomposition based on the phase space reconstruction to analyze the bearing vibration signal and then utilized CNN to process the obtained features for bearing fault diagnosis. Zhu et al. [104] proposed a symmetrized dot pattern to transform vibration signals into 2-D images and the trained a convolutional network for fault diagnosis. Li et al. [105] used the K-singular value decomposition to enhance the resolution of time-frequency features obtained by Wigner-Ville Distribution and then built CNN for planetary gearbox fault classification. Udmale et al. [106] utilized the kurtogram of raw signals to train CNN for bearing fault diagnosis. In [107], Senanayaka proposed a gearbox fault diagnosis method based on multiple classifiers and data fusion. Specifically, the vibration spectrum was used as the input of multilayer perceptron while the features from STFT and CWT were used input of CNN. Finally, the naïve Bayes combiner was employed to integrate the results of two classifiers.

In this subsection, the applications of 2-D CNN for fault classification were reviewed systematically. A clear and intuitive summary is displayed in Table 10, which aims to help reader search these studies quickly depending on the signal transform approach or analysis object.

Table 10. Summary of the applications of 2-D CNN for fault classification.

| Transform method | References | Object |
|---|---|---|
| Data matrix | [41] [42] [43] [44] [45] [46] | Bearing |
| | [47] [48] / [49] / [50] | Gear/bearing, wind turbine/bearing, pump |
| | [51] / [52] | Motor/compressor |
| Images | [53] [54] [55] [56] | Bearing |
| | [57] [58] [59] / [60] | Bearing, gear/bearing, pump |
| Time or frequency domain transform | [62] [63] [64] [65] [66] [67] [68] | Bearing/bearing, gear/gear |
| | [69] / [61] / [70] [71] [72] | |
| Wavelet transform | [73] [74] [75] [76] [77] | Bearing |
| | [78] [79] [80] [81] / [82] | Gear/gear, rotor, bearing |
| | [83] [84] / [85] / [86] / [87] | Rotor/motor/bearing, gear/tool |
| Short time Fourier transform | [88] [89] / [90] [91] | Bearing/bearing, gear |
| | [92] / [93] | Bearing, pump/ motor |
| Other | [94] [95] [97] [98] [99] [100] | Bearing/gear |
| | [103] [106] / [101] [102] [105] | |
| | [96] / [104] / [107] | Rotor/ bearing, rotor/ bearing, rotor, gear |

**4.1.2. 1-D convolutional network based classification**

In addition to 2-D convolutional network based fault classification, a more direct strategy is to construct 1-D convolutional diagnostic model to process original time-series data. In this subsection, the applications of 1-D convolutional network for fault classification are reviewed according to raw sensor data types, such as vibration and AE data.

- Vibration data

Vibration data has been the most common source of information in machine fault diagnosis due to its legibility and intuitiveness, including convolutional network based intelligent diagnosis. Eren [108] used raw vibration signal as the input to train 1-D CNN for IMS bearing fault detection. Pan et al. [109] combined the CNN with the long short term memory (LSTM) network and proposed an improved bearing fault diagnosis method. Inspired by the second generation wavelet transform, Pan et al. [110] improved the convolutional network and proposed a LiftingNet to process raw mechanical data for fault classification. Qian et al. [111] constructed an adaptive overlapping CNN for bearing fault diagnosis, in which the raw vibration signals were used to train model. Jia et al. [112] proposed a normalized CNN for imbalanced fault classification of machinery. In this method, a neuron activation maximization algorithm was presented to help understand the feature learning process of the network. Eren et al. [113] developed a compact adaptive 1-D CNN for real-time bearing fault diagnosis. Ma et al. [114] integrated the residual convolutional network, deep belief network as well as deep auto-encoder and proposed an ensemble deep learning method for fault diagnosis of rotor bearing system. Wang et al. [115] proposed a multi-scale learning network with the 1-D and 2-D convolution channels to learn the local correlation of adjacent and nonadjacent intervals in vibration signals for bearing fault diagnosis. Huang et al. [116] added a multi-scale cascade layer at the front of conventional convolutional network and proposed a CNN approach with multi-scale information for bearing fault diagnosis. Qiao et al. [117] used raw vibration signals as the input and proposed an adaptive weighted multiscale CNN for bearing fault diagnosis under variable operating conditions. Abdeljaber et al. [118] utilized the compact CNN to present an online condition monitoring method for fault detection and severity identification of bearings.

Huang et al. [119] utilized raw vibration signals as the model input and proposed a deep decoupling CNN for intelligent compound fault diagnosis. Liu et al. [120] combined the denoising convolutional autoencoder with CNN to develop an anti-noise fault diagnosis method. Han et al. [121] proposed an enhanced convolutional network with enlarged receptive fields for planetary gearbox fault diagnosis. Considering the inherent multiscale characteristics of vibration signals, Jiang et al. [122] developed a multi-scale CNN for wind turbine gearbox fault diagnosis. In [123], Sun et al. firstly used back-propagation based neural network to learn the local filters. Then these local filters were used to build the feed-forward convolutional neural network for feature learning. Finally, the learned features were fed into SVM classifier for induction motor fault classification. Yuan et al. [124] applied multi-sourced heterogeneous monitoring data as the input and presented a multi-mode CNN based method for rotor system fault diagnosis. Afrasiabi et al. [125] proposed an accelerated CNN for bearing fault diagnosis of induction motors, in which the pruning connection and weight sharing technique were used to compress model without loss of accuracy. Chen et al. [126] utilized 1-D CNN to learn features from raw vibration signals and then fed these features into a bidirectional LSTM network for wear state identification of tool.

In addition to directly using the raw vibration data, some scholars adopted the features extracted from vibration data to train 1-D convolutional neural network for fault diagnosis. Xie et al. [127] used the convolutional network to learn features from frequency spectrum of vibration signals and then integrated these features with energy entropy of empirical mode decomposition and time domain features for final fault classification. Sadoughi et al. [128] used spectral kurtosis and envelope spectrum analysis to process raw mechanical data and proposed a physics-based CNN for fault diagnosis of rotating machinery. To maintain the diagnosis performance in the noisy environment and different working load, Zhang et al. [129] developed a CNN with training interference based diagnosis method. In [130], Dong et al. used 1-D CNN and 2-D CNN to respectively extract features from the frequency spectrum and STFT spectrum of vibration signals for rolling bearing degradation. Jing et al. [131] developed a CNN based method to extract features from the frequency spectrum of vibration signal for gearbox fault diagnosis. In [132], Ma et al. used coefficients of wavelet packet decomposition as the model input and proposed a lighted CNN for bearing fault diagnosis.

- Other data

Compared with vibration data, other mechanical data are also used in 1-D convolutional network based classification, including current, AE, and build-in encoder data. For instance, Ince et al. [133] directly used raw current signals to train 1-D CNN for real-time motor condition monitoring, the results show the effectiveness and superiority of their approach. Besides, they proposed a real-time broken rotor bar fault detection model based on the shallow 1-D convolutional neural network [134]. Khan et al. [135] used the motor current as the input and developed an analytical model for inter-turn fault diagnosis by combining the 1-D CNN and LSTM network. Kao et al. [136] applied a 1-D CNN to the fault diagnosis of magnet synchronous motor over a wide speed range by using current data as the model input. In [137], the motor vibration signal and the stator current signal were firstly segmented by analysis windows of varying lengths for the joint representation. Then, the CNN and LSTM network were designed to automatically learn discriminative features and achieve motor fault diagnosis. In addition to the current data, AE data analysis is also a common monitoring manner for fault diagnosis. Li et al. [138] used convolutional network and gate recurrent unit to reprehensively extract features from AE and vibration data. Then the learned features were concatenated for gear pitting fault diagnosis. In [139], Appana et al. utilized the CNN to process the envelope spectrums of AE signals to achieve bearing fault diagnosis under varying rotating speeds. In light of the drawbacks of external sensors, Jiao et al. [14] proposed a build-in encoder information based CNN for intelligent fault diagnosis. In this method, a multivariate encoder information was presented by information fusion to capture comprehensive mechanical health states, then the convolutional network was designed for adaptively feature learning and condition classification.

According to the types of sensor data, the comprehensive review on the applications of 1-D convolutional network is presented in this subsection. Depending on above literature, a concise generalization is displayed in Table 11.

Table 11. Summary of the applications of 1-D CNN for fault classification.

| Signal Type | References | Object |
| --- | --- | --- |
| Vibration data | [108] [109] [110] [111] [112] [113] [114] [115] [116] [117] [118] | Bearing |
|  | [119] [120] [121] | Bearing, gear |
|  | [122] / [123] [124] [125] / [126] | Wind turbine gearbox/motor/tool |
|  | [127] [128] [129] [130] [132] / [131] | Bearing/gear |
| Other data | [133] [134] [135] [137] / [136] [139] / [138] [14] | Motor/bearing/gear |

### 4.1.3. Classification based on convolutional network variants

As introduced in Section 3.2, many variants of CNN have also been studied and applied in the field of fault diagnosis. Thus in this subsection, we will review these publications according to different network variants.

- Applications of ResNet to fault classification

Zhao et al. [140] employed a series of wavelet packet coefficients as the model input and proposed a deep residual network with dynamically weighted wavelet coefficients for planetary gearbox fault diagnosis under serious noise environment. The comparison results showed higher accuracies than other deep learning approaches. Furthermore, they [141] combined the WT with ResNet and proposed the multiple wavelet coefficients fusion based deep residual network for planetary gearbox fault diagnosis, which aimed to learn more easily-distinguished features from the input data. Li et al. [142] proposed a deep residual learning network for fault diagnosis, in which the data augmentation techniques were presented to artificially create additional valid samples for model training. The result showed that their method can achieve high diagnosis accuracy with small original training dataset. Zhang et al. [143] used raw vibration signals as the model input to train a deep ResNet for bearing fault diagnosis, the results show the superiority to traditional CNN model. Peng et al. [144] presented a deeper 1-D CNN with residual learning for fault diagnosis of wheelset bearings in high-speed trains. Ma et al. [145] proposed a deep residual convolutional network based on a separable convolution and concatenated ReLU lightweight convolution for bearing fault diagnosis, in which the coefficients of wavelet packet transform were selected as the network input. Zhuang et al. [146] proposed a stacked residual dilated CNN for bearing fault diagnosis by combining the dilated convolution, the input gate structure of LSTM and the residual network. Su et al. [147] utilized raw time sequences as the input and presented a residual-squeeze net for fault diagnosis of high-speed train bogie. Ma et al. [148] proposed a fault diagnosis method of planetary gearbox under nonstationary running conditions using ResNet with demodulated time-frequency features. Considering the non-stationary conditions of machine, Liu et al. [149] proposed multi-scale kernel based residual convolutional network for motor fault diagnosis. The results showed the superiority compared with state-of-the-art methods. From above review, it can be seen that ResNet diagnostic model is promising for more comprehensive feature extraction and higher diagnosis accuracy conditions in modern industry, especially in complicated mechanical equipment or industrial environment.

- Applications of GAN to fault classification

Due to the excellent data generation characteristics, GANs have been gradually applied to the field of fault diagnosis, especially for the diagnostic scenario with imbalanced data sets. Cao et al. [150] firstly transformed the time-domain signals into image data. Then a GAN was designed for rolling bearing fault classification. The results illustrated the potential of GAN on the fault diagnosis with small samples. Xie et al. [151] developed a GAN to generate the samples of minority classes for bearing fault diagnosis, which aimed to address the issue of data imbalance. Shao et al. [152] proposed an auxiliary classifier GAN based diagnostic framework to generate synthesized data and achieve induction motor fault diagnosis. Afrasiabi et al. [153] combined GAN with temporal CNN and proposed a wind turbine fault diagnosis method, in which the former was used as the feature extractor and the latter was used as the fault classifier. Li et al. [154] proposed an enhanced GAN for fault diagnosis of rotating machinery with imbalanced data. In their method, 2-D convolutional network was used to build the generator and discriminator, which aims to produce small samples to balance the dataset. Suh et al. [155] employed the nested scatter plot method to transform raw vibration signals into 2-D images, then a 2-D CNN was designed for bearing fault classification. In addition, a GAN was embedded in this framework to generate fault images for the data imbalance issue. To address the issue of lacking the labeled fault data, Guo et al. [156] proposed a multi-label 1-D GAN for fault diagnosis, in which the auxiliary classifier GAN was used to generate real damage data and then the generated data and real data are both used to train fault classifier. The experimental results showed that the proposed method can improve diagnosing accuracy from 95% to 98% when model was trained with the generated data.

- Applications on other variants to fault classification

Jiao et al. [9] employed the built-in encoder and external vibration signals as the input in parallel and presented a deep coupled dense convolutional network based intelligent fault diagnosis. The results verified the superiority than traditional convolutional network. Li et al. [157] presented an improved inception network to process raw vibration signals for gear pitting fault diagnosis. In [158], Chen et al. proposed a deep inception net with atrous convolution to bridge the gap between artificial and real damage for bearing fault diagnosis. Zhu et al. [159] employed the STFT to convert signals into 2-D graphs as the input and proposed a capsule network with an inception block and a regression branch for bearing fault diagnosis. Chen et al. [160] proposed a deep capsule network with stochastic delta rule for rolling bearing fault diagnosis, in which raw vibration signals were used as the model input.

**4.1.4. Summary**

This subsection reviews the classification applications in machine fault diagnosis using convolutional neural networks, in which the results reveal that powerful feature learning and fault identification capabilities of CNN. Despite these methods have achieved certain success, some existing practical problems still cannot be ignored. For instance, the success of convolutional neural networks is based on the large scale datasets with a tremendous amount of labeled samples. However, in many practical situations, a large number of labeled samples are inaccessible, especially for fault data. Besides, above most approaches assume that the distributions of training data and test data are same, however, this assumption is not hold in real industry. Consequently, it is necessary to solve these realistic problems and advance convolutional diagnostic approaches for the promising employment in modern intelligent industry.

**4.2. Applications on health prediction**

Different from the fault classification, the purpose of health prediction is to track the degraded state of machinery, even if no apparent failure has occurred. This branch is vitally important in the field of machine fault diagnosis, which allows maintenance personnel make early judgments and decisions to avoid losses and injuries. Therefore, the applications on health prediction are reviewed and summarized according to the application object in this section.

**4.2.1. Health predication of bearing**

Rolling bearings are widely used and play an important role in modern machinery. The deterioration or failure of bearings will lead to machine breakdown and even disaster. Therefore, numerous studies have been conducted to assess and predict the health condition of bearings. Yoo et al. [161] used the continuous WT to obtain the time-frequency images for the health indicator (HI) construction. Then the CNN was designed to process these images for bearing RUL prediction. Belmiloud et al. [162] used wavelet packet decomposition to extract features as the model input and then presented a deep CNN based method for adaptive HI construction. Hinchi et al. [163] proposed a bearing RUL estimation method, in which the convolutional layer and LSTM layer were integrated to learned features from raw sensor data. Guo et al. [164] proposed a method for HI construction, in which the trend burr was considered and the results showed their proposed method is more effective than other methods in terms of tradability, monotonicity and scale similarity. Ren et al. [165] presented a spectrum-principal-energy-vector algorithm to obtain the eigenvector as the network input to train the CNN for bearing RUL prediction. She et al. [166] proposed a multi-channel CNN with exponentially decaying learning rate to construct wear indicator and evaluate the health of rolling bearing, in which the original multi-channel signals were used as the input. Mao et al. [167] employed the CNN to learn features from the marginal spectrum of Hilbert-Huang transform. After that, a LSTM network was constructed for RUL prediction of bearings. Li et al. [168] used the STFT to process raw vibration signals to obtain the time-frequency domain information. Then a deep CNN was built to extract multi-scale features for RUL estimation. Zhu et al. [169] used the WT to acquire the time-frequency representation and then trained a multi-scale CNN to learn global and local features for RUL estimation. The results showed enhanced performance in prediction accuracy compared to tradition data-driven and CNN based methods. Wang et al. [170] converted 1-D signals into the 2-D images to train CNN for RUL prediction, in which the maximum correlation entropy with regular terms was employed as the loss function for better performance compared to the mean square error. Zhang et al. [171] proposed a deep multilayer perceptron convolutional network for HI construction, in which the outlier region correction method is introduced to detect and remove outliers and enhance the interpretability of HI. Yang et al. [172] utilized raw mechanical signals to trained a double-CNN model for RUL prediction, in which the first CNN was used to identify the incipient fault point and the second CNN model was applied for RUL prediction. Considering that the uncertainty was critical for health prognostic, Peng et al. [173] introduced a Bayesian multi-scale convolutional network based prognostic method, which shows the more accurate performance than point estimates. Yao et al. [174] combined the empirical model decomposition with ensemble CNNs for bearing RUL estimation, in which the former can reveal the nonstationary property of degradation data and help CNN to get a more accurate prediction. Wang et al. [175] integrated the CNN and LSTM network to process time-series data and calculate an unsupervised $H$-statistic for bearing performance degradation assessment. Liu et al. [176] presented a joint-loss CNN for bearing fault recognition and RUL prediction in parallel, which can capture common features between different relative tasks and improve the generalization capability. Wang et al. [177] proposed a deep separable convolutional network for RUL prediction of machinery, in which the data from different sensors were used to train a separable convolutional building block with a residual connection for feature learning. They further proposed a recurrent convolutional network for RUL prediction [178], in which recurrent convolutional layers were designed to model the temporal dependencies and variational inference was utilized to quantify the uncertainty of prediction results.

**4.2.2. Health predication of turbofan engine**

Benefitting from the public C-MAPSS dataset, many researchers validated the proposed prognosis approaches on the turbofan engine. Babu et al. [179] constructed 2-D data matrix from multi-variate time series to train a CNN with two-convolution layers and two-fully connected layers for RUL estimation. Li et al. [180] adopted the time window approach to process the multi-variate temporal data and then developed a deep CNN for feature extraction and RUL estimation. Wen et al. [181] presented a deep residual CNN for RUL estimation, in which the *k*-fold ensemble method was adopted to enhance the prediction preformation. Li et al. [182] presented a directed acyclic graph network for RUL prediction by combining CNN and LSTM network. The comparative results showed that the proposed method had better predication accuracy. Al-Dulaimi et al. [183] proposed a hybrid deep network framework for RUL estimation, in which the LSTM network and CNN were arranged in parallel for feature learning and then a multilayer fully connected network was designed for feature fusion and decision making. Ruiz-Tagle Palazuelos et al. [184] introduced a capsule neural network for degradation estimation of turbofan engine and the results showed the superiority than traditional CNN based methods. Kong et al. [185] adopted the polynomial regression to construct HI from raw data and then designed a hybrid deep model based on the CNN and LSTM network for RUL prediction.

**4.2.3. Other health predication application**

In addition to the bearing and turbofan engine, some scholars also developed convolutional network based prediction approaches to other applications, such as CNC machine. Zhao et al. [186] proposed a convolutional bi-directional long short term memory network for CNC machining tool health monitoring, which combined the advantages of CNN and LSTM network to obtain accurate predication. Qiao et al. [187] proposed a hybrid deep learning framework for gearbox fault diagnosis and tool wear prediction, in which the multiple convolutional and LSTM layers were firstly designed to extract local spatiotemporal features and then a holistic convolution-LSTM layer was designed to extract holistic spatiotemporal features. Aghazadeh et al. [188] employed CNN to establish a deep learning algorithm for tool wear estimation, in which wavelet transform and spectral subtraction algorithms were designed to intensify the effect of tool wear and reduce the effect of cutting parameters. Huang et al. [189] utilized features from time-domain, frequency domain and time-frequency domain of multi-sensor signals as health information and proposed a deep CNN based method for tool wear prediction. In [190], Fu et al. combined the CNN and LSTM network to establish the logical relationship of observed variables for condition monitoring of wind turbine gearbox bearing. Kong et al. [191] presented a health monitoring method of wind turbines based SCADA data. In their approach, the CNN and gated recurrent units were integrated to learn spatial and temporal features, and then the exponential weighted moving average control chart was designed for condition recognition. Luo et al. [192] employed the dual-tree complex wavelet to obtain multiscale characteristics as the input. Then an enhanced convolutional LSTM network was designed for damage monitoring of the automotive suspension component. Li et al. [193] developed a scalable degradation assessment approach for bandsaw machine by proposing a dual-phase modeling method. In this approach, a physics informed model is firstly established to generate the HI to monitor wear condition using the vibration and acoustic signals. Then a deep CNN based surrogate model is designed to replace the physics informed model by using alternative low-cost sensor data.

**4.2.4. Summary**

In this section, applications of convolutional networks on health prediction are systemically reviewed according to the application object. The summarization reveals that many researchers have successfully developed convolutional networks based prediction approaches to address weak generality, flexibility and intelligence of previous physical and mathematical models. However, it should be noticed that lifetime data is difficult to obtain in practical industry. In other words, there is usually no sufficient data to train a complete life prediction model, hence how to build a model with experimental or simulation data and then make the model generalizable to practical industrial applications should be paid to more attention.

**4.3. Applications on transfer diagnosis**

Although CNN on fault classification and health predication of machinery have acquired certain achievements, an assumption that training data and test data have same data distribution is necessary for most of the above approaches. In practical industrial scenarios, the data distribution differences are inevitable due to natural wear of equipment, changes in operating conditions, interference from environment and human, and so on. Consequently, the performance of above most models will seriously degrade when the data distributions between training set (source domain) and test set (target domain) are different. An immediate solution is to retrain or built new model, however, a large number of labeled data are necessary in this case. In many task scenarios, sufficient labeled instances are either difficult to collect, or their labeling costs is prohibitively. Therefore, it is quite necessary to explore how to apply the previously models established on the related domain to the new diagnostic scenarios. In light of these issues, transfer learning or domain adaptation technologies have been introduced to machine fault diagnosis, especially its combination with the deep convolutional networks. In this section, a summary on the applications of convolutional network on transfer diagnosis will be introduced in detail. Before starting the literature review, three common tricks are firstly introduced, including parameters transfer, moment matching strategies, and adversarial domain adaptation.

**4.3.1. Parameters transfer**

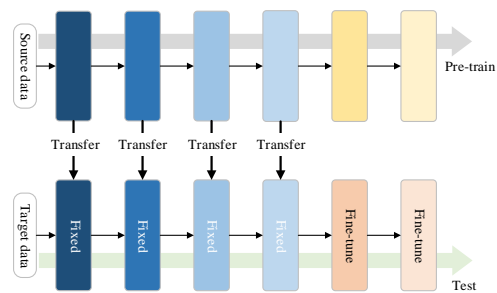

Fig. 18. Illustration of parameters transfer.

Parameters transfer is also called pre-train model based transfer, which means that the partial parameters of network trained in the source domain are fixed and transferred while remaining parameters will be fine-tuned using labeled data in the target domain. Intuitive understanding is shown in Fig. 18, in which the idea comes from the fact that the features of previous layers are general and transferable while the features of the last few layers are task-specific. Therefore, the model can be used to new target domain by using the few labeled target data to fine-tune parameters of task-specific layers.

Some researchers have applied this technology to achieve transfer diagnosis of machinery in recent years. Cao et al. [194] proposed a deep CNN based transfer learning approach for gearbox fault diagnosis. The first part of their method was constructed by a part of a pre-trained network and the second part was a fully connected layers retrained by gear data. Hasan et al. [195] used the frequency spectrum as the input and presented a parameters transfer based 1-D convolutional network for bearing fault diagnosis under variable working conditions. Hemmer et al. [196] employed a CNN pre-trained by ImageNet dataset to learn features from WT images of vibration and AE signals. Then a sparse autoencoder-based SVM was designed to process these features for bearing fault classification. Zhong et al. [197] proposed a transfer learning framework for gas turbine fault diagnosis, in which the CNN trained on large-scale annotated normal dataset was transferred to fault diagnosis task with limited fault data for feature learning and then the SVM was designed as the new classifier for fault classification. Wen et al. [198] converted the raw time-domain signals to RGB images to fine-tune a pre-trained ResNet-50 for fault diagnosis. Furthermore, they utilized the negative correlation learning to retrain several fully-connected layers and the Softmax classifier of the pre-trained ResNet-50 for fault classification [199]. Han et al. [200] presented a transfer learning framework for fault diagnosis of unseen machine conditions, in which the CNN trained on large datasets was transferred to new tasks with proper fine-tuning. In addition, they designed three transfer learning strategies to investigate the feature transferability in the different network levels. Shao et al. [201] employed the WT to convert raw signals into images and developed a deep transfer framework for machine fault diagnosis, in which the labeled time-frequency images were used to fine-tune the higher layers of a convolutional network pre-trained by ImageNet dataset. Ma et al. [202] introduced the frequency slice wavelet transform to extract the raw vibration signals into 2-D time-frequency images and then fine-tune a pre-trained AlexNet model for bearing fault diagnosis. Hasan et al. [203] used the acoustic spectral images of AE signals to reflect mechanical health state and then proposed a pre-train CNN based parameters transfer learning approach for bearing fault diagnosis under variable speed conditions. Chen et al. [204] employed the parameters transfer based method for fault diagnosis of rotary machinery, in which a wide kernel 1-D CNN was designed for learning transferable features and results showed the effectiveness of the proposed method.

In this subsection, applications on parameters transfer based fault diagnosis are reviewed. Although this transfer strategy is easy to understand and operate, a distressing issue still exists that labeled data in the target domain is necessary. Therefore, these transfer learning approaches will encounter unexpected obstacles due to the unavailability of labeled data in practical industrial applications.

### 4.3.2. Discrepancy measure based transfer diagnosis

Discrepancy measure based method is generally achieved by minimizing a certain distance between hidden activations of convolutional network in different domains, in which the key is to explore the efficient discrepancy metric function, such as Maximum Mean Discrepancy (MMD) and correlation alignment. In this subsection, the definition of MMD and correlation alignment is firstly introduced to help understand this transfer learning manner, then the applications on discrepancy measure based transfer diagnosis are systemically reviewed.

Maximum mean discrepancy [205, 206] measures the distribution divergences by the mean embedding of two distributions in the reproducing kernel Hilbert space $\mathcal{H}$. Specifically, give the datasets $\mathcal{D}_s = \{\mathbf{x}_i^s\}$ and $\mathcal{D}_t = \{\mathbf{x}_i^t\}$ drawn from two domains with different distributions $P$ and $Q$, the MMD can be calculated as:

$$\mathcal{L}_{MMD} = \sup_{\phi \in \mathcal{H}} (\mathbb{E}_P[\phi(\mathbf{x}^s)] - \mathbb{E}_Q[\phi(\mathbf{x}^t)]) \tag{7}$$

where $\phi$ represents the feature map; $P = Q$ if and only if $\mathcal{L}_{MMD} = 0$. In practical application, the MMD is calculated as the empirical estimation based on the kernel mean embedding:

$$\hat{\mathcal{L}}_{MMD} = \left\| \frac{1}{n_s} \sum_{i=1}^{n_s} \phi(\mathbf{x}_i^s) - \frac{1}{n_t} \sum_{i=1}^{n_t} \phi(\mathbf{x}_i^t) \right\|_{\mathcal{H}}^2 = \frac{1}{n_s^2} \sum_{i=1}^{n_s} \sum_{j=1}^{n_s} k(\mathbf{x}_i^s, \mathbf{x}_j^s) + \frac{1}{n_t^2} \sum_{i=1}^{n_t} \sum_{j=1}^{n_t} k(\mathbf{x}_i^t, \mathbf{x}_j^t) - \frac{2}{n_s n_t} \sum_{i=1}^{n_s} \sum_{j=1}^{n_t} k(\mathbf{x}_i^s, \mathbf{x}_j^t)$$

$$(8)$$

where $k(\cdot,\cdot)$ represents the characteristic kernel; $n_s$ and $n_t$ are the number of source samples and target samples.

Correlation alignment [207] is defined as a constraint function to measure the data distribution difference based on the second-order statistics. Mathematically, given the source feature matrix $D_s \in \mathbb{R}^{n_s \times d}$ and target feature matrix $D_t \in \mathbb{R}^{n_t \times d}$, where the row represents the number of sample and the column denotes the feature dimension. The covariance matrices of two feature matrices can be calculated as:

$$\begin{aligned} C_s &= \frac{1}{n_s - 1}(D_s^\mathrm{T} D_s - \frac{1}{n_s}(\mathbf{1}^\mathrm{T} D_s)^\mathrm{T}(\mathbf{1}^\mathrm{T} D_s)) \\ C_t &= \frac{1}{n_t - 1}(D_t^\mathrm{T} D_t - \frac{1}{n_t}(\mathbf{1}^\mathrm{T} D_t)^\mathrm{T}(\mathbf{1}^\mathrm{T} D_t)) \end{aligned} \tag{9}$$

where $C_s$ and $C_t$ stand for the covariance matrices of two domains, respectively. $\mathbf{1}$ is a column vector with all elements equal to 1; $\cdot^\mathrm{T}$ stands for the transposition. Based on two covariance matrices, the correlation alignment is defined as follows:

$$\mathcal{L}_c = \frac{1}{4d^2} \left\| C_s - C_t \right\|_F^2 \tag{10}$$

where $\|\cdot\|_F^2$ represents the squared matrix Frobenius norm.

In the community of fault diagnosis, Zhang et al. [208] proposed a domain adaptation based CNN for fault diagnosis under varying working conditions, in which the MMD was used to minimize the domain divergence. Li et al. [209] presented a CNN based rolling bearing fault diagnosis method under noisy and changing working condition. In their approach, a feature clustering method was introduced to minimize the difference of intra-class and maximize the difference of inter-class. Meanwhile, the MMD was adopted to reduce the domain divergences. Furthermore, they proposed a multi-layer domain adaptation approach for bearing fault diagnosis [210], in which the multi-kernel maximum mean discrepancy was employed as the metric function to reduced distribution differences between different domains. In [211], Xiao et al. presented a domain adaptation based motor fault diagnosis method, in which the CNN was adopted to extract multi-level features from raw vibration signals and the MMD was incorporated to reduce the feature distribution differences of multiple layers. Yang et al. [212] combined the CNN with MMD to introduce a transfer learning network for the fault diagnosis from laboratory bearings to locomotive bearings. Han et al. [213] extended the marginal distribution adaptation to the joint distribution adaptation and proposed a deep transfer network for fault diagnosis. Xu et al. [214] proposed a convolutional transfer feature discrimination network for unbalanced fault diagnosis under variable rotational speeds, in which the MMD was used to reduce the distribution differences of high-dimensional features. Zhu et al. [215] converted raw vibration data into gray pixel images as the network input and proposed a multi-Gaussian kernels MMD based deep transfer learning approach for rolling bearing fault diagnosis under different operating conditions. To address the cross-domain diagnosis problem with insufficient target samples, Li et al. [216] used the MMD based generative convolutional networks to generate fake target fault samples and proposed a domain adaptation based approach for bearing fault diagnosis. In [217], a renewable fusion method was proposed for fault diagnosis under variable speed conditions and unbalanced samples, in which the second order statistics were used to reduce feature distribution differences and the contrastive loss function was employed to promote the similar features between different speeds.

### 4.3.3. Adversarial learning based transfer diagnosis

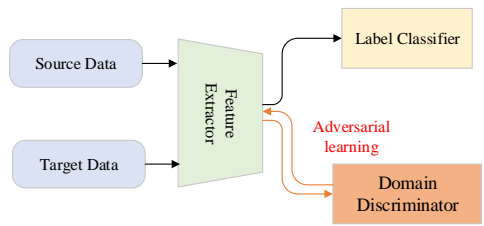

Fig. 19. Illustration of adversarial domain adaptation.

In addition to above two moment matching algorithms, another transfer learning strategy, named the adversarial domain adaptation [218], is attracting increasing attentions. Unlike the discrepancy measure based method, adversarial domain adaptation constructs a two-player minimax game to supervise the network for learning transferable features. Specifically, adversarial domain adaptation network is usually composed of a feature extractor $F$, a label classifier $C$ and a domain discriminator $D$ as shown in Fig. 19, in which the discriminator is trained to distinguish whether the features are from the source domain or the target domain while the feature extractor tries to learn the features and fool the discriminator. In the training process, meanwhile, the classifier is trained to minimize the classification error of source data. Give the source domain $\mathcal{D}_s = \{(\mathbf{x}_i^s, y_i^s)\}_{i=1}^{n_s}$ with $n_s$ samples and target domain $\mathcal{D}_t = \{(\mathbf{x}_i^t)\}_{i=1}^{n_t}$ with $n_t$ samples. The overall objective of the adversarial domain adaptation network is described as follows:

$$\mathcal{L}_{ada} = \frac{1}{n_s}\sum_{\mathbf{x}_i \in \mathcal{D}_s} \mathcal{L}_y(C(F(\mathbf{x}_i)), y_i) - \frac{\lambda}{n_s + n_t}\sum_{\mathbf{x}_i \in (\mathcal{D}_s \cup \mathcal{D}_t)} \mathcal{L}_d(D(F(\mathbf{x}_i)), d_i) \quad (11)$$

where $\mathcal{L}_y$ represents the classification loss function; $\mathcal{L}_d$ denotes the domain identification loss function; $y_i$ and $d_i$ stand for the category label and domain label, respectively; $\lambda$ is the trade-off parameter.

Inspired by the adversarial domain adaptation, Han et al. [219] introduced a deep adversarial convolutional network for machine fault diagnosis and the results showed that the proposed method was superior to the conventional convolutional networks. Guo et al. [220] integrated the moment matching with adversarial learning strategies to develop a deep convolutional transfer learning network for fault diagnosis of machines, in which the training and test dataset were acquired from different machines. Zhang et al. [221] proposed a Wasserstein distance guided multi-adversarial convolutional network for fault diagnosis under different operating conditions. The experimental results showed improved performance than MMD based methods. To improve the Wasserstein distance-based adversarial approach, Wang et al. [222] presented a triplet loss guided adversarial domain adaptation network for bearing fault diagnosis and the results showed the better performance. Xie et al. [223] proposed a transfer learning approach for fault diagnosis using the cycle-consistent GAN, in which the GAN was designed to generate new sample for unknown conditions to pre-train a classifier. From the perspective of decision boundaries, Jiao et al. [224] developed an unsupervised adversarial adaptation network to achieve cross-domain fault diagnosis using two task classifiers without the domain discriminator. Furthermore, they presented a domain adaptation network based on classifier inconsistency for addressing more realistic problem [225], i.e. partial transfer diagnosis, in which the source domain and target domain have different class number.

In addition to above three popular transfer diagnosis methods, several ingenious technologies for transfer diagnosis are also introduced. For example, Zhang et al. [226] used raw vibration signals as the model input and presented a CNN based diagnosis method, in which the wide convolutional kernels and adaptive batch normalization were adopted for the domain adaptation capability. Duan et al. [227] proposed an auxiliary model based domain adaptation method for reciprocating compressor diagnosis under different operating conditions, in which a pre-trained CNN was used for feature learning and a marginalized stacked auto-encoder was used to eliminate data distribution difference. Hasan et al. [228] introduced a discrete orthonormal Stockwell transform for data preprocessing. Then a deep convolutional network was proposed for bearing fault diagnosis under variable rotational speeds. Xiao et al. [229]

proposed a transfer learning model for fault diagnosis by integrating the modified TrAdaBoost algorithm with the convolutional network.

**4.3.4. Summary**

In this section, applications on transfer diagnosis are methodically reviewed and a concise summary is listed in Table 12. The first column represents different transfer strategies and the third column denotes specific methodology for achieving transfer diagnosis. The last column stands for the application scenarios, where "image→machine" represents that the pre-trained model is from the field of image processing; "operating conditions" and "machines" denotes the transfer between different conditions or different machines, respectively.

Table 12. Summary of applications on transfer diagnosis.

| Transfer strategy | References | Methodology | Scenarios |
| --- | --- | --- | --- |
| Parameters transfer | [194] [196] [198] [199] [201] [202] | Pre-train by image data | image→machine |
| | [195] [197] [200] [203] | Pre-train by mechanical data | operating conditions |
| | [204] | | operating conditions and machines |
| Discrepancy measure | [208] [209] [210] [211] [213] [214] [215] [216] | MMD | operating conditions |
| | [212] | | machines |
| | [217] | Correlation alignment | operating conditions |
| Adversarial transfer | [219] | Adversarial discriminator | operating conditions |
| | [220] | discriminator and MMD | machines |
| | [221] [222] | Adversarial Wasserstein | operating conditions |
| | [223] | GAN | operating conditions |
| | [224] [225] | Classifier discrepancy | operating conditions |
| Other | [226][227] [228] [229] | / | operating conditions |

## 5. Conclusions

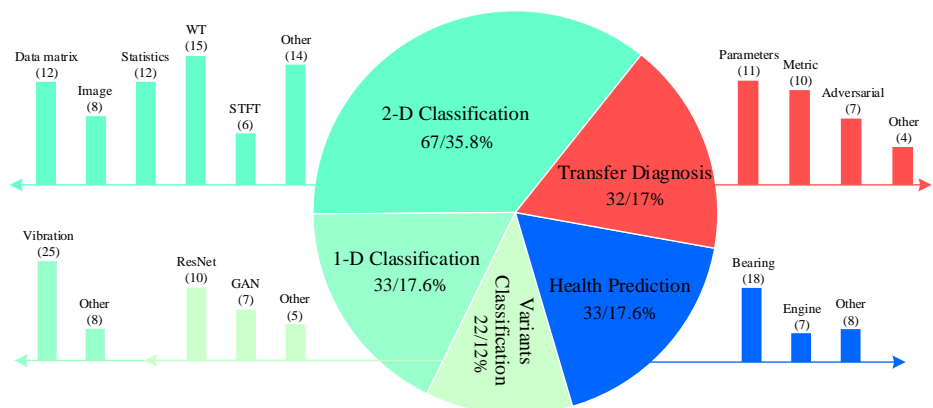

Fig. 20. The pie chart of publication related to the convolutional network based monitoring diagnosis method.

In previous sections, the published literature on CNFD has been systemically reviewed. The overall pie chart of these summarizations is displayed in Fig. 20. It should be pointed out that the related literature on this field is huge and abundant, meanwhile the new research is also emerging. Therefore, it is inevitable that some papers are missing from the current review. In addition, some non-English publications are also not considered in this work because of the limitation of language proficiency. In spite of this, some observations and conclusions are provided in this section based on the current literature review.

(1) From Fig. 20, it can be found that more than 65% of publications are focused on the fault classification task. And most of transfer diagnosis approaches are oriented to the health condition classification. The number of mechanical health prognostic applications only account for about 17.6% of the total. This phenomenon may be because the implementation of fault classification is more easy and intuitive while the health prognostic usually requires additional assistance, such as the HI construction and health stage division etc. However, the fault classification mainly focuses on various failure conditions, which are the final states of machine. Prior to these conditions, the equipment usually undergoes a degradation process, in which the foreseeable action should be taken instead of waiting for the final fault. Thus the health prediction, i.e. degradation monitoring or RUL prediction, should be paid more attentions in future studies.

(2) Almost all models mentioned in the above literature are trained and tested in the experimental or simulated scenario, thus these models may be unsuitable to be directly applied to realistic industry since the acquired data and industrial data usually have certain differences. In addition, the training data and test data even come from the same set of experiment in some validation process in some research, in which the excellent results can be produced owing to the data similarity. This will make researchers blindly believe in the network capability. Consequently, utilizing the reasonable experimental data and diverse realistic industrial data to train more powerful models is of significance.

(3) It is known that the parameter set plays a decisive role in the performance of deep convolutional network. Reviewing the above literature, the design and selection of network parameters (including architecture and hyper-parameters) are mainly determined by authors' subjectivity while a specific standard or rule for selecting appropriate parameters has not been formed. Although a set of parameters cannot be ideally applied to various tasks, the study on relation between parameters and mechanical signal characteristics or parameter selection trick is still promising and significant.

(4) In some applications, especially 2-D convolutional networks, the data transform or signal processing technology is necessary. Thus it will increase the complexity of the overall framework and reduce the efficiency as well as the level of intelligence. On the contrary, raw signals based applications can avoid the requirements for domain knowledge and construct the end-to-end diagnostic framework. But the noise and interference existed in raw data may disturb the model convergence and even lead the model astray. Therefore, it is suggested to objectively view the merit and demerit between the deep convolutional network and advanced signal processing algorithms, and organically integrate them for better performance.

## 6. Prospects

Despite the published literature has achieved great advancements in machine fault diagnosis, there still several aspects need to be further explored and investigated. Therefore, in this section, we will share some prospects with the readers, researchers and engineers who aim to promote the development of this field.

(1) More theoretical investigation is necessary to reveal the "black box" issue of CNFD approaches. Although convolutional networks have been widely applied to fault diagnosis, in-depth theoretical research is still very rare. For example, the relation between the weights and the mechanical features or the explanation of the learned features have not been reasonably explained. In addition, the "black box" issue will make the companies or factories doubt the capabilities of these methods and refuse to apply them to realistic scenarios. Therefore, it is urgent to lift the veil of CNFD methods whether in academia or industry.

(2) How to identify unseen damage types or fault conditions?

The literature on the applications of CNDP generally only focuses on identifying the faults existed in the training set. It means that the model can be used only when the category of the test dataset is included in the training set. However, the construction of all-encompassing training set is expensive or even impossible. Moreover, some strange faults will inevitably occur in real scenarios with the changing of equipment itself and working environment. As a result, it is still an open question to explore the models which could distinguish the unseen damages or faults.

(3) There is a requirement to detect early damage from the point of quantitative analysis.

Most of CNFD applications only focus on how to identify the different health categories, however, there is no obvious fault types in early degradation stage. In particular, it is unreasonable to carry out diagnosis until the occurrence of large or significant failures in high-precise and vital industrial applications. Moreover, the decision should be performed according to the level of damage by the qualitative analysis. Considering this issue, therefore, the effort to explore the detection and quantitative analysis of early weak damages should be encouraged.

(4) How to train the model using non-stationary data for diagnosis or prognosis under variable operating condition?

In previous variable operating conditions studies, the convolutional models are usually trained by the data of smooth operation or the time-invariant features extracted by signal processing technologies. As a result, the former is only from one stationary condition to another and the later need certain expert knowledge. Moreover, collecting the stationary data is difficult and even impossible in the realistic continuously non-stationary operating environment. Consequently, how to use the deep convolutional network to directly process the non-stationary data and achieve the reliable diagnosis is also an urgent problem.

(5) How to speed up the CNFD algorithm for the real-time diagnostic requirements?

It is known that a frequently mentioned drawback for the CNFD approaches is that they consume more time for training than classical shallow algorithms, thus this will lead to unsatisfactory in real-time and quick task requirement. Therefore, the exploration of the novel technology and trick to accelerate and improve the CNFD algorithms requires to be paid more attentions in future research.

(6) How to establish the CNFD models that can be used for the requirement of equipment fleet?

According to the above literature, existing methods are mostly employed for the diagnosis and prognosis of a single machine. However, the cluster machine development is becoming an increasing trend in the rapid manufacturing and production era. Therefore, it is more significant to study the CNFD models with powerful generalization capability which can be freely applied to the other similar machines.

(7) How to utilize the opportunity of industrial big data to improve the performance of CNDP methods?

The sufficient data is the premise and foundation to achieve the excellent performance of deep networks. Reviewing the literature, the choice of data quantity heavily depends on the subjective factors or is limited by the experimental condition. As a result, many models are not optimized to optimal performance due to the illusion of simple or limited amount data. Therefore, how to seize the chance of industrial big data and utilize its characteristic, such as diversity and heterogeneity, to develop more robust and reliable models will be another promising topic in next research.

**Acknowledgments**

This research was supported by National Natural Science Foundation of China (Grant No. 51421004, 91860205), the Defense Industrial Technology Development Program (Grant No. JCKY2018601C013), Fundamental Research Funds for the Central Universities (Grant No. xzy022019022) and the China Scholarship Council.

bearings, IEEE Trans. Ind. Electron., 62 (3) (2014) 1781-1790.

[30] http://www.mfpt.org/FaultData/FaultData.htm.

[31] https://drive.google.com/open id=1_ycmG46PARiykt82ShfnFfyQsaXv3_VK.

[32] https://figshare.com/articles/Gear_Fault_Data/6127874/1.

[33] V. Nair, G.E. Hinton. Rectified linear units improve restricted boltzmann machines. In: Editor edito. Pub Place; 2010. p. 807-814.

[34] A.L. Maas, A.Y. Hannun, A.Y. Ng. Rectifier nonlinearities improve neural network acoustic models. In: Editor edito. Pub Place; 2013. p. 3.

[35] K. Jarrett, K. Kavukcuoglu, M. Ranzato, Y. LeCun. What is the best multi-stage architecture for object recognition? In: Editor edito. Pub Place: IEEE; 2009. p. 2146-2153.

[36] S. Ioffe, C. Szegedy, Batch normalization: Accelerating deep network training by reducing internal covariate shift, arXiv preprint arXiv:1502.03167, (2015).

[37] N. Srivastava, G. Hinton, A. Krizhevsky, I. Sutskever, R. Salakhutdinov, Dropout: a simple way to prevent neural networks from overfitting, The journal of machine learning research, 15 (1) (2014) 1929-1958.

[38] K. He, X. Zhang, S. Ren, J. Sun. Deep residual learning for image recognition. In: Editor edito. Pub Place; 2016. p. 770-778.

[39] G. Huang, Z. Liu, L. Van Der Maaten, K.Q. Weinberger. Densely connected convolutional networks. In: Editor edito. Pub Place; 2017. p. 4700-4708.

[40] I. Goodfellow, J. Pouget-Abadie, M. Mirza, B. Xu, D. Warde-Farley, S. Ozair, A. Courville, Y. Bengio. Generative adversarial nets. In: Editor edito. Pub Place; 2014. p. 2672-2680.

[41] X. Guo, L. Chen, C. Shen, Hierarchical adaptive deep convolution neural network and its application to bearing fault diagnosis, Measurement, 93 (2016) 490-502.

[42] W. Fuan, J. Hongkai, S. Haidong, D. Wenjing, W. Shuaipeng, An adaptive deep convolutional neural network for rolling bearing fault diagnosis, Meas. Sci. Technol., 28 (9) (2017) 95005.

[43] H. Shao, H. Jiang, H. Zhang, W. Duan, T. Liang, S. Wu, Rolling bearing fault feature learning using improved convolutional deep belief network with compressed sensing, Mech. Syst. Signal Pr., 100 (2018) 743-765.

[44] H. Shao, H. Jiang, H. Zhang, T. Liang, Electric Locomotive Bearing Fault Diagnosis Using a Novel Convolutional Deep Belief Network, IEEE Trans. Ind. Electron., 65 (3) (2018) 2727-2736.

[45] S. Wang, J. Xiang, Y. Zhong, Y. Zhou, Convolutional neural network-based hidden Markov models for rolling element bearing fault identification, Knowl-Based Syst., 144 (2018) 65-76.

[46] W. Gong, H. Chen, Z. Zhang, M. Zhang, R. Wang, C. Guan, Q. Wang, A Novel Deep Learning Method for Intelligent Fault Diagnosis of Rotating Machinery Based on Improved CNN-SVM and Multichannel Data Fusion, Sensors-Basel, 19 (7) (2019) 1693.

[47] L. Jing, T. Wang, M. Zhao, P. Wang, An adaptive multi-sensor data fusion method based on deep convolutional neural networks for fault diagnosis of planetary gearbox, Sensors-Basel, 17 (2) (2017) 414.

[48] H. Chen, N. Hu, Z. Cheng, L. Zhang, Y. Zhang, A deep convolutional neural network based fusion method of two-direction vibration signal data for health state identification of planetary gearboxes, Measurement, 146 (2019) 268-278.

[49] T. Han, C. Liu, L. Wu, S. Sarkar, D. Jiang, An adaptive spatiotemporal feature learning approach for fault diagnosis in complex systems, Mech. Syst. Signal Pr., 117 (2019) 170-187.

[50] Y. Yang, H. Zheng, Y. Li, M. Xu, Y. Chen, A fault diagnosis scheme for rotating machinery using hierarchical symbolic analysis and convolutional neural network, ISA Trans., (2019).

[51] R. Liu, G. Meng, B. Yang, C. Sun, X. Chen, Dislocated time series convolutional neural architecture: An intelligent fault diagnosis approach for electric machine, IEEE Trans. Ind. Inform., 13 (3) (2016) 1310-1320.40